\documentclass[12pt]{article}
\usepackage{amssymb,amscd,array}
\catcode `\@=11
\@addtoreset{equation}{subsection}

\def\harr#1#2{\smash{\mathop{\hbox to .5in{\rightarrowfill}}
\limits^{\scriptstyle#1}_{\scriptstyle#2}}}

\def\harrl#1#2{\smash{\mathop{\hbox to .5in{\leftarrowfill}}
\limits^{\scriptstyle#1}_{\scriptstyle#2}}}

\def\qed{\blacksquare}
\newcommand{\be}{\begin{equation}}
\newcommand{\ee}{\end{equation}}
\newcommand{\bea}{\begin{eqnarray}}
\newcommand{\eea}{\end{eqnarray}}
\newcommand{\R}{\mathbb{R}}
\newcommand{\N}{\mathbb{N}}
\newcommand{\C}{\mathbb{C}}
\newcommand{\Z}{\mathbb{Z}}
\newtheorem{thm}{Theorem}[section]
\newtheorem{rem}[thm]{Remark}

\newtheorem{cor}[thm]{Corollary}
\newtheorem{prop}[thm]{Proposition}
\begin{document}
\begin{titlepage}
%\thispagestyle{empty}
%\begin{flushright}
%IFA-FT-402-1994, November
%\end{flushright}
%\bigskip\bigskip
\begin{center}
{\bf \Large{Ward Identities and Renormalization \\of General Gauge Theories}}
\end{center}
\vskip 1.0truecm
\centerline{D. R. Grigore
\footnote{e-mail: grigore@theor1.theory.nipne.ro, grigore@ifin.nipne.ro}}
\vskip5mm
\centerline{Dept. of Theor. Phys., Inst. Atomic Phys.}
\centerline{Bucharest-M\u agurele, P. O. Box MG 6, ROM\^ANIA}
\vskip 2cm
\bigskip \nopagebreak
\begin{abstract}
\noindent
We introduce the concept of general gauge theory which includes Yang-Mills
models. In the framework of the causal approach and show that the anomalies can
appear only in the vacuum sector of the identities obtained from the gauge
invariance condition by applying derivatives with respect to the basic fields.
Then we provide a general result about the absence of anomalies in higher
orders of perturbation theory. This result reduces the renormalizability proof
to the study of lower orders of perturbation theory. For the Yang-Mills model
one can perform this computation explicitly and obtains its renormalizability
in all orders.
\end{abstract}
%\newpage\setcounter{page}1
\end{titlepage}

\section{Introduction}

The causal approach to perturbative renormalization theory of by Epstein and
Glaser \cite{EG1}, \cite{Gl} gives significant simplification of the conceptual
and computational aspects for quantum electrodynamics \cite{Sc1}, \cite{DF1},
\cite{qed}, Yang-Mills theories \cite{DHKS1}, \cite{DHKS2}, \cite{DHS2},
\cite{DHS3}, \cite{AS}, \cite{ASD3}, \cite{Du1}, \cite{Du2}, \cite{Hu1}, \cite
{Kr1}, \cite{DS}, \cite{YM}, \cite{standard}, \cite{fermi}, \cite{Sc4},
\cite{Sc5}, \cite{ren-ym}, gravity \cite{Gri1}, \cite{Gri2}, \cite{SW}, the
analysis of scale invariance can be done \cite{scale}, \cite{Pr3}, Wess-Zumino
model \cite{wz}, etc. In this approach one uses exclusively the Bogoliubov
axioms of renormalization theory \cite{BS2} imposed on the scattering matrix:
this is an operator acting in the Hilbert space of the model, which is a Fock
space generated from the vacuum by the quantum fields corresponding to the
particles of the model. If one considers the $S$-matrix as a perturbative
expansion in the coupling constant of the theory, one can translate these
axioms on the chronological products.  Epstein-Glaser approach is a inductive
procedure to construct the chronological products in higher orders starting
from the first-order of the perturbation theory - the interaction Lagrangian -
which is a Wick polynomial. For gauge theories one can construct a non-trivial
interaction only if one considers a larger Hilbert space generated by the
fields associated with the particles of the model and the ghost fields.  The
condition of gauge invariance becomes in this framework the condition of
factorization of the $S$-matrix to the physical Hilbert space in the adiabatic
limit. To avoid infra-red problems one works with a formulation of this
factorization condition which corresponds to a formal adiabatic limit and it is
perfectly rigorously defined \cite{DHKS2}. The obstructions to the
implementation of the condition of gauge invariance are called anomalies.  The
most famous is the Adler-Bell-Bardeen-Jackiw anomaly (see \cite{Ni1} for a
review). The most convenient way to organize the combinatorial argument seems
to be the following one \cite{Sto1}, \cite{DF1}.  One constructs the
chronological products $ T(W_{1}(x_{1}),\dots,W_{n}(x_{n})) $ associated to
arbitrary Wick monomials $ W_{1}(x_{1}),\dots,W_{n}(x_{n}) $ according to
Epstein-Glaser prescription \cite{EG1} (which reduces the induction procedure
to a distribution splitting of some distributions with causal support) or
according to Stora prescription \cite{PS} (which reduces the renormalization
procedure to the process of extension of distributions).

If
$T(x)$
is the interaction Lagrangian (i.e. the first order chronological product) and
$d_{Q}$
the BRST operator, we suppose the validity of some ``descent" equations of the
type:
\bea
d_{Q} T(x) = i \partial_{\mu}T^{\mu}(x), \quad
d_{Q} T^{\mu}(x) = i \partial_{\nu}T^{\mu\nu}(x), \dots,
\nonumber \\
d_{Q} T^{\mu_{1},\dots,\mu_{p-1}}(x) = 
i \partial_{\mu_{p}}T^{\mu_{1},\dots,\mu_{p}}(x), \quad
d_{Q} T^{\mu_{1},\dots,\mu_{p}}(x) = 0
\label{descent}
\eea
for some finite $p$. One denotes by
$A^{k}(x),~ k = 1,2,\dots,$
the expressions
$T(x),~T^{\mu}(x),~T^{\mu\nu},\dots$
and we suppose that these expressions have a well defines ghost number, i.e all
terms of the Wick polynomial
$A^{k}(x)$
have the same ghost degree. Then we can write the preceding equation in the
compact form
\be
d_{Q} A^{k}(x) = i \sum_{m} c^{k;\mu}_{m}
{\partial\over \partial x^{\mu}} A^{m}(x), \quad k = 1, 2,\dots
\ee
for some constants
$c^{k;\mu}_{m}$.
The gauge invariance condition has the generic form
\be
d_{Q} T(A^{k_{1}}(x_{1}),\dots,A^{k_{n}}(x_{n})) = 
i \sum_{l=1}^{n} (-1)^{s_{l}} \sum_{m} c^{k_{l};\mu}_{m}
{\partial\over \partial x^{\mu}_{l}}
T(A^{k_{1}}(x_{1}),\dots,A^{m}(x_{l}),\dots,A^{k_{n}}(x_{n}))
\label{gauge-model}
\ee
for all 
$n \in \N$
and all
$k_{1}, \dots, k_{p} = 1,2,\dots$.
Here the expression
\be
s_{l} \equiv \sum_{i=1}^{l-1} gh(A^{k_{i}}).
\ee

One can define \cite{DF1} the notion of derivative of a Wick polynomial with
respect to the basic fields of the model. In particular one can apply these
derivatives operators to the polynomials
$A^{k}(x),~ k = 1,2,\dots,$.

Then we can prove that the gauge invariance condition can be reduced to some
identities verified by the vacuum expectation values of the chronological
products of the following type:
$
<\Omega, T(DA^{k_{1}}(x_{1}),\dots,DA^{k_{n}}(x_{n})) \Omega>;
$
here
$DA^{k}(x),~ k = 1,2,\dots$
are polynomials in the derivatives of the basic expressions
$A^{k}(x),~ k = 1,2,\dots.$
These are the so-called C-g identities in the language of \cite{DHKS1} -
\cite{DHS2};  it is plausible to expect that they are equivalent to the Ward
(Slavnov - Taylor) identities from the usual formulation of gauge theories so
we prefer to call them Ward identities.

This idea was used in \cite{DF1} to study the conservation of the
electromagnetic current in quantum electrodynamics. The generalization of these
idea to non-Abelian gauge theories is under current investigation \cite{BDF}.

We can show that the anomalies of these Ward identities are absent in higher
orders of the perturbation theory (more precisely in orders greater than the
dimension of the space-time $+1$). In this way, one can check the gauge
invariance of the Yang-Mills model if one establishes that there are no
anomalies in the orders $2, 3, 4$ and $5$.

We start in the next Section with a systematic study of the Wick monomials. In
particular, we circumvent the complications associated with the signs coming
from the fields with Fermi-Dirac statistics using Grassmann variables
(following a suggestion from \cite{Sc1}). In Section \ref{pert} we sketch the
framework of the perturbative renormalization theory of Bogoliubov. In the next
Section we formulate the notion of general gauge theory and we derive the Ward
identities. In the last Section we check the absence of anomalies in lower
orders of the perturbation theory for the Yang-Mills model.
\newpage

\section{The General Framework\label{framework}}

\subsection{Free Fields\label{free}}

We define here the general framework of a free field theory in the Fock space
following closely the point of view of \cite{DF1}. Some standard notions from
quantum relativistic mechanics are used \cite{Va}, \cite{EG1}. We define a 
{\it system of free fields} to be the ensemble
$(\phi^{A}(x), {\cal F}, \Omega, U_{a,L})$
where:

(i) $\phi^{A}(x), \quad A = 1,\dots,N$
are distribution-valued operators acting in the Fock space ${\cal F}$ with a
common dense domain $D_{0}$. Here
$x \in M$
where $M$ is the Minkowski space.

(ii) $\Omega \in D_{0}$ is called the {\it vacuum state}. The vectors
$\phi^{A_{1}}(x_{1}) \dots \phi^{A_{n}}(x_{n}) \Omega$
generate the Fock space ${\cal F}$.

(iii) $a,L \mapsto U_{a,L}$ is a unitary representation of the group
$SL(2,\C)$
acting in ${\cal F}$
such that
\be
U_{a,L} \phi^{A}(x) U_{a,L}^{-1} 
= {S(L^{-1})^{A}}_{B} \phi^{B}(\delta(L)\cdot x + a);
\ee
here 
$SL(2,\C) \ni L \mapsto \delta(L) \in {\cal L}^{\uparrow}_{+}$
is the covering map and
$SL(2,\C) \ni L \mapsto S(L)$
is a $N \times N$ representation of $SL(2,\C)$.

(iv) $supp(\widetilde{\psi^{A}}) \subset V^{+}_{M_{A}} \cup V^{-}_{M_{A}}$
where
$M_{A} \geq 0$
is called the {\it mass} of the field $\psi^{A}, \quad A = 1,\dots,N$.

(v) Let us denote by
$\phi^{A}_{\pm}$
the {\it positive (negative) frequency components} of $\psi^{A}$
such that
$supp(\widetilde{\phi^{A}_{\pm}}) \subset V^{\pm}_{M_{A}}$;
then there exist a system of numbers
$z_{A} \in \Z, \quad A= 1,\dots,N$
such that the following {\it canonical (anti)commutation relations} are true:
\bea
\phi^{A}_{\pm}(x) \phi^{B}_{\pm}(y) 
= (-1)^{z_{A} z_{B}} \phi^{B}_{\pm}(y) \phi^{A}_{\pm}(x),
\nonumber \\
\phi^{A}_{\pm}(x) \phi^{B}_{\mp}(y) 
- (-1)^{z_{A} z_{B}} \phi^{B}_{\mp}(y) \phi^{A}_{\pm}(x)
= D^{AB}_{\pm}(x-y) \times {\bf 1}_{\cal F}
\label{CCR}
\eea
where
$D^{AB}_{\pm}(x)$
are distributions verifying:
\be
D^{AB}_{\pm}(x) = 0, \quad iff \quad  z_{A} + z_{B} \not= 0,
\ee
and if we define
\be
D^{AB}(x) \equiv D^{AB}_{+}(x) + D^{AB}_{-}(x)
\ee
then these distributions have causal support:
$supp(D^{AB}(x)) \subset V^{+} \cup V^{-}$.

(vi) One has
\be
\phi^{A}_{-}(x) \Omega = 0, \quad \forall A = 1,\dots,N.
\ee

(vii) {\it Equations of motion} of the type
\be
\sum_{\alpha} u^{\alpha}_{A} \partial_{\alpha} \phi^{A}(x) = 0, 
\quad A = 1,\dots,N 
\label{motion}
\ee
for some constants
$u^{\alpha}_{A}$
are verified; here we use Schwartz multi-indices $\alpha, \beta, \dots$
but one can also use alternative notation
$u^{\mu\nu\dots}_{A}$
from jet-bundle extension theory.  One cannot avoid the existence of the
equations of motion: indeed, because of the requirement (iv) the fields will
verify Klein-Gordon equation:
\be
\partial^{2} \phi^{A} + M_{A}^{2} \phi^{A} = 0, \quad A = 1,\dots, N
\label{KG}
\ee
i.e. the preceding equation is true for
\be
u^{\mu\nu}_{A} = M_{A}^{2} g^{\mu\nu} u_{A}
\label{kg}
\ee
for arbitrary numbers 
$u_{A}$.
For Dirac fields one has a first order system of equations of motion: the Dirac
equation.

(viii) For some of the fields some reality condition might be imposed,
connecting the Hermitian conjugates
$(\phi^{A})^{*}$
and the original fields
$\phi^{B}$.

Let us remind the fact that from (ii) one can derive that if the (graded)
commutators of some operator $X$ with all fields $\phi^{A}$ are zero, then this
operator is proportional to the unit operator 
${\bf 1}$
from the Fock space.

One can easily see that all known models in Fock spaces can be accommodated in
this scheme. 

We avoid the complications due to the signs from (\ref{CCR}) if we consider a
$\Z$-graduated Grassmann algebra 
$
{\cal G} = \sum_{n \in \Z} {\cal G}_{n}
$
over $\C$ and some Grassmann numbers
$g_{A} \in {\cal G}$
which are invertible and of parity $z_{A}, \quad \forall A = 1,\dots,N$.
Then we consider distributions with values in
${\cal G} \otimes {\cal L}({\cal F})$
(here
${\cal L}({\cal F})$
are the linear operators from ${\cal F}$)
given by:
\be
\varphi^{A}(x) \equiv g_{A} \otimes \phi^{A}(x), \quad \forall A = 1,\dots,N.
\ee

We call these operators the {\it supersymmetric associated fields}. We consider
$J^{r}(\R^{N}, M)$
the r-th order jet bundle extension of the trivial fibre bundle
$\R^{N} \times M \rightarrow M$
which describes classical fields with $N$ components defined over the Minkowski
space 
$M \sim \R^{4}$
and we consider the jet bundle coordinates
$u^{\alpha}_{A}, \quad A = 1, \dots, N, \quad |\alpha| \leq r$. The natural
number $r$ should be chosen large enough. Then we define the operators:
\be
\varphi^{\pm}_{u}(x) \equiv
\sum_{\alpha,A} u^{\alpha}_{A} \partial_{\alpha} \varphi^{A}_{\pm}(x)
\ee
and
\be
\varphi_{u}(x) \equiv \varphi^{+}_{u}(x) + \varphi^{-}_{u}(x).
\ee

We call {\it mass-shell} the linear subspace:
\be
{\cal M} \equiv \{ u \in J^{r}(\R^{N}, M) \vert \varphi_{u} = 0 \}
\ee
and we denote
$[u] \equiv u \quad {\rm modulo} \quad {\cal M}$. We see that in fact the
operators
$\varphi_{u}(x)$
depend only on the equivalence class $[u]$ i.e. we can consistently use the
notation:
\be
\varphi_{[u]}(x) \equiv \varphi_{u}(x).
\ee

One can verify elementary that we have the following form of the canonical
commutator relations:
\bea
[\varphi^{\pm}_{u}(x), \varphi^{\pm}_{w}(y) ] = 0,
\nonumber \\
~[\varphi^{\pm}_{u}(x), \varphi^{\mp}_{w}(y) ] = \Delta^{\pm}_{uw}(x-y),
\nonumber \\
~[\varphi_{u}(x), \varphi_{w}(y) ] = \Delta_{uw}(x-y)
\label{CCR-Grassmann}
\eea
where in the left hand side we have the {\bf usual} commutator and we have
defined:
\bea
\Delta^{\pm}_{uw}(x-y) \equiv \sum_{\alpha,A} \sum_{\beta,B}
g_{A} g_{B} \quad u^{\alpha}_{A} w^{\beta}_{B} 
\partial^{x}_{\alpha} \partial^{y}_{\beta}D^{AB}_{\pm}(x-y),
\nonumber \\
\Delta_{uw}(x-y) \equiv \Delta^{+}_{uw}(x-y) + \Delta^{-}_{uw}(x-y).
\eea

One can easily see that the distribution
$\Delta_{uw}(x-y)$
has causal support and we have the symmetry property
\be
\Delta_{uw}(x-y) = \Delta_{wu}(y-x).
\ee

It is convenient to choose the Grassmann algebra ${\cal G}$ such that
${\cal G}_{0} = \C$;
then we have
$\Delta^{\pm}_{uw}(x-y) \in \C$.

The fact that in (\ref{CCR-Grassmann}) we have the ordinary commutator will
make all the following computations more convenient because will we not worry
about the Jordan signs appearing for fields with Fermi-Dirac statistics. 

We will need in the following a derivative operation defined on the classical
fields jet bundle: if $\alpha$ is multi-index, then we define
$\partial_{\alpha}: J^{r}(\R^{N}, M) \rightarrow J^{r}(\R^{N}, M)$
according to:
\be
(\partial_{\alpha}u)^{\beta}_{A} \equiv \sum_{\alpha+\beta=\gamma}
u^{\gamma}_{A}.
\ee

Then we have three elementary facts:
\begin{prop}
(i) If 
$u \in {\cal M}$ 
then
$\partial_{\alpha}u \in {\cal M}$
for any multi-index $\alpha$.
\end{prop}
{\bf Proof:}
One applies to the equations of motion (\ref{motion}) the partial derivative
operator
$\partial_{\alpha}$.
$\qed$
\begin{prop}
The following relations are true:
\be
\partial_{\alpha}\varphi_{u}(x) = \varphi_{\partial_{\alpha}u}(x),
\ee
\be
\partial_{\alpha}\Delta_{uw} = \Delta_{\partial_{\alpha}u,w}.
\ee
\end{prop}
{\bf Proof:}
The first relation is a result of a elementary computation. For the second
relation, we apply the partial derivative operator
$\partial_{\alpha}$
to the last canonical commutation relation (\ref{CCR-Grassmann}) and use the
first relation.
$\qed$

We also note that there is a natural group action
$l, u \mapsto l\cdot u$
of the group
$SL(2,\C)$
on the elements of
$J^{r}(\R^{N}, M).$
The Hermitian conjugation operation postulated at item (viii) of the preceding
Subsection induces a natural conjugation operation
$u \mapsto u^{*}$
on the classical fields from
$J^{r}(\R^{N}, M).$
We will need these operations later.

We end this Subsection pointing the fact that in the literature one usually
uses Grassmann valued classical fields instead of the classical fields from
$J^{r}(\R^{N}, M)$.
The connection is
$J^{r}(\R^{N}, M) \ni u^{\alpha}_{A} \rightarrow g_{A} u^{\alpha}_{A}
\in J^{r}({\cal G} \otimes \R^{N}, M)$.

The classical structure
$J^{r}(\R^{N}, M)$
associated to the Wick monomials algebra was used somewhat differently in
\cite{Bo} and \cite{DF1}.

\subsection{Supersymmetric Wick Monomials\label{s-wick}}

A complete and rigorous investigation of the Wick combinatorial arguments can
be found in \cite{DG1}. Here we give an approach which does not use Feynman
graphs. We use consistently Bourbaki conventions
$
\sum_{\emptyset} \equiv 0, \quad \prod_{\emptyset} \equiv 1.
$
We will define Wick monomials through the following proposition:
\begin{prop}
The operator-valued distributions
$N(\varphi_{u_{1}}(x_{1}),\dots,\varphi_{u_{n}}(x_{n}))$
are uniquely determined through the following properties:
\be
N(\varphi_{u_{1}}(x_{1}),\dots,\varphi_{u_{n}}(x_{n})) \Omega
= \varphi^{+}_{u_{1}}(x_{1}),\dots,\varphi^{+}_{u_{n}}(x_{n}) \Omega;
\ee
\be
[ N(\varphi_{u_{1}}(x_{1}),\dots,\varphi_{u_{n}}(x_{n})) , \varphi_{w}(y) ]
= \sum_{l=1}^{n} 
N(\varphi_{u_{1}}(x_{1}),\dots,\widehat{\varphi_{u_{l}}},\dots,
\varphi_{u_{n}}(x_{n})) \Delta_{u_{l}w}(x_{l}-y);
\ee
\be
N(\emptyset) \equiv {\bf 1}.
\ee

In the first two relations $n$ is arbitrary.
\label{normal}
\end{prop}
{\bf Proof}: Is is elementary. For $n = 1$ we find from the second property
that
$N(\varphi_{u}(x)) - \varphi_{u}(x)$
commutes with every operator
$\varphi_{w}(y)$
so it must be a of the form
${\rm const} \times {\bf 1}.$ 
But the first relation fixes this constant to $0$. Next, we suppose that we
have defined the expressions
$N(\varphi_{u_{1}}(x_{1}),\dots,\varphi_{u_{n-1}}(x_{n-1}))$
and we use the second and the first relation to define the action of
$N(\varphi_{u_{1}}(x_{1}),\dots,\varphi_{u_{n}}(x_{n}))$
on vectors of the type
$\varphi_{w_{1}}(y_{1}),\dots,\varphi_{w_{k}}(y_{k})\Omega$;
from (ii) of the previous Subsection we know that they generate the whole Fock
space.  
$\qed$

We call the operators
$N(\varphi_{u_{1}}(x_{1}),\dots,\varphi_{u_{n}}(x_{n}))$
{\it supersymmetric Wick} (or {\it normal}) {\it monomials in $n$ variables}. 

Let us note that in fact, the Wick monomial 
$N(\varphi_{u_{1}}(x_{1}),\dots,\varphi_{u_{n}}(x_{n}))$
depends only on the equivalence classes
$[u_{1}],\dots,[u_{n}]$.
Using induction, one can easily prove that it is completely symmetric in the
arguments. 

We can easily establish the connection with the usual definition of the Wick
monomials:
\begin{prop}
The following relation is true:
\be
N(\varphi_{u_{1}}(x_{1}),\dots,\varphi_{u_{n}}(x_{n}))
= \sum_{I,J \in {Part\{1,\dots,n\}}}
\prod_{i \in I} \varphi^{+}_{u_{i}}(x_{i}) \quad 
\prod_{j \in J} \varphi^{-}_{u_{j}}(x_{j}).
\ee
\label{normal-pm}
\end{prop}
{\bf Proof:}
We first note that the order of the factors in the two products is irrelevant
because of the commutativity property (\ref{CCR-Grassmann}). The proof consists
in denoting the right hand side of the relation by
$
N^{\prime}(\varphi_{u_{1}}(x_{1}),\dots,\varphi_{u_{n}}(x_{n}))
$
and proving the the three relations appearing in the preceding proposition are
true. Then we use the uniqueness assertion.
$\qed$

As a immediate corollary we obtain:
\begin{cor}
The following relations are true:
\bea
N(\varphi_{u_{1}}(x_{1}),\dots,\varphi_{u_{n}}(x_{n}),\varphi_{w}(y))
= N(\varphi_{u_{1}}(x_{1}),\dots,\varphi_{u_{n}}(x_{n})) \varphi_{w}(y)
\nonumber \\
- \sum_{l=1}^{n} <\Omega,\varphi_{u_{l}}(x_{l}) \varphi_{w}(y) \Omega>
N(\varphi_{u_{1}}(x_{1}),\dots,\widehat{\varphi_{u_{l}}},\dots,
\varphi_{u_{n}}(x_{n}));
\eea
\bea
N(\varphi_{u_{1}}(x_{1}),\dots,\varphi_{u_{n}}(x_{n}),\varphi_{w}(y))
= \varphi_{w}(y) N(\varphi_{u_{1}}(x_{1}),\dots,\varphi_{u_{n}}(x_{n})) 
\nonumber \\
+ \sum_{l=1}^{n} \Delta^{+}_{u_{l}w}(x_{l}-y) 
N(\varphi_{u_{1}}(x_{1}),\dots,\widehat{\varphi_{u_{l}}},\dots,
\varphi_{u_{n}}(x_{n}))
\eea
\label{normal1}
\end{cor}
{\bf Proof:}
The first relation follows immediately from the preceding proposition. If we
combine it with the second property from Prop. \ref{normal} the we obtain the
second relation.
$\qed$

If we apply the preceding results we can also obtain using induction
\begin{cor}
The normal products can be expressed as
\be
N(\varphi_{u_{1}}(x_{1}),\dots,\varphi_{u_{n}}(x_{n}))
= \sum_{k} \sum_{i_{1} < \cdots < i_{k}}
d^{+}(x_{1},\dots,x_{n})
\varphi_{u_{i_{1}}}(x_{i_{1}}),\dots,\varphi_{u_{i_{k}}}(x_{i_{k}})
\ee
where
$d^{+}(x_{1},\dots,x_{n})$
are distributions.
\end{cor}

In fact, one can express the distributions from the statement as sum of
distributions
$d^{+}_{G}$
labelled by Feynman graphs \cite{DG1}. However, we do not need this result here.

Now, a non-trivial observation is that if we formally ``colapse" all arguments
$x_{1},\dots,x_{n} \mapsto x$
in the expression
$
N(\varphi_{u_{1}}(x_{1}),\dots,\varphi_{u_{n}}(x_{n}))
$
we obtain well-defined operators:
\begin{prop}
The expressions
\be
W_{u_{1},\dots,u_{n}}(x) \equiv
N(\varphi_{u_{1}}(x),\dots,\varphi_{u_{n}}(x))
\ee
are well defined and they are completely symmetric in the indices
$u_{1},\dots,u_{n}$.
\label{wick-def}
\end{prop}
{\bf Proof:}
We collapse the arguments in the relations of the preceding proposition and
formally obtain:
\be
W_{u_{1},\dots,u_{n}}(x) \Omega
= \varphi^{+}_{u_{1}}(x),\dots,\varphi^{+}_{u_{n}}(x) \Omega;
\ee
\be
[ W_{u_{1},\dots,u_{n}}(x) , \varphi_{w}(y) ]
= \sum_{l=1}^{n} 
W_{u_{1},\dots,\widehat{u_{l}},\dots,u_{n}}(x) \Delta_{u_{l}w}(x_{l}-y);
\ee
\be
W_{\emptyset} \equiv {\bf 1}.
\ee

The only non-trivial step is to prove that the right hand side of the first
relation is well defined; this is done in \cite{WG}. The rest of the proof is
identical.
$\qed$

If
$U \equiv \{u_{1},\dots,u_{n}\}$
then we can use consistently the notation
$W_{U}(x)$.
We call expressions of this type a {\it supersymmetric Wick monomials of rank}
$n$
{\it in one variable}. Again we note that the dependence on the classical
fields
$u_{1},\dots,u_{n}$
is only through the equivalence classes. The action
$l, u \mapsto l\cdot u$
of the group
$SL(2,\C)$
on the jet bundle coordinates extends naturally to the action
$l, U \mapsto l\cdot U$
componentwise. The same assertion is valid for the hermitian conjugation
$u \mapsto u^{*}$
which extends componentwise to
$U \mapsto U^{*}$.
We now give an elementary result:
\begin{prop}
The following formula is true:
\be
\partial_{\alpha}W_{u_{1},\dots,u_{n}}(x) =
\sum_{l=1}^{n} W_{u_{1},\dots,\partial_{\alpha}u_{l},\dots,u_{n}}(x).
\ee
\end{prop}
{\bf Proof:}
It follows by induction commuting both sides with an arbitrary field
$\varphi_{w}(y)$.
$\qed$

By definition, a {\it (supersymmetric) Wick polynomial} is any linear
combination (with coefficients from ${\cal G}$) of Wick monomials. The set of
all (supersymmetric) Wick polynomials in the Fock space ${\cal F}$ is denoted
by 
${\rm sWick}({\cal F})$. 
The action of
$SL(2,\C)$
and the Hermitian conjugation extend naturally to the set of (supersymmetric)
Wick polynomials. When no ambiguity is possible we abandon the attribute
supersymmetric.

A generalization of the collapsing procedure used above is available and
essential to the perturbation theory. Namely, we consider the expression
$N(\varphi_{u_{1}}(x_{1}),\dots,\varphi_{u_{k}}(x_{k}))$
for $k > n$ and group the variables
$x_{1},\dots,x_{k}$
in $n$ subsets; then we collapse the arguments to distinct points inside every
subset.
\begin{prop}
The expressions
\be
N(W_{U_{1}}(x_{1}),\dots,W_{U_{n}}(x_{n})) \equiv
N(\prod_{u \in U_{1}} \varphi_{u}(x_{1}),\dots,
\prod_{u \in U_{n}} \varphi_{u}(x_{n})).
\ee
are well-defined and completely symmetric in the arguments.
\label{multi-normal}
\end{prop}
{\bf Proof:}
As before, we obtain from the first proposition the following relations:
\be
N(W_{U_{1}}(x_{1}),\dots,W_{U_{n}}(x_{n})) \Omega
= \prod_{i=1}^{n} \prod_{u \in U_{i}} \varphi^{+}_{u}(x_{i}) \Omega;
\ee
\bea
[ N(W_{U_{1}}(x_{1}),\dots,W_{U_{n}}(x_{n})) , \varphi_{w}(y) ]
\nonumber \\
= \sum_{l=1}^{n} \sum_{u \in U_{l}}
N(W_{U_{1}}(x_{1}),\dots,W_{U_{l}-\{u\}},\dots,W_{U_{n}}(x_{n})) 
\Delta_{uw}(x_{l}-y);
\label{normal-type}
\eea
\be
N(W(x)) \equiv W(x)
\ee
and we can use recursion.
$\qed$

We have results similar to Corollary \ref{normal1}:
\begin{cor}
The following relations are true:
\bea
N(W_{U_{1}}(x_{1}),\dots,W_{U_{k}}(x_{k}),\varphi_{w}(y))
= N(W_{U_{1}}(x_{1}),\dots,W_{U_{k}}(x_{k})) \varphi_{w}(y)
\nonumber \\
- \sum_{l=1}^{k} \sum_{u \in U_{l}}
<\Omega,\varphi_{u}(x_{l}) \varphi_{w}(y) \Omega>
N(W_{U_{1}}(x_{1}),\dots,W_{U_{l}-\{u\}},\dots,W_{U_{k}}(x_{k}));
\eea
\bea
N(W_{U_{1}}(x_{1}),\dots,W_{U_{k}}(x_{k}),\varphi_{w}(y))
= \varphi_{w}(y) N(W_{U_{1}}(x_{1}),\dots,W_{U_{k}}(x_{k})) 
\nonumber \\
+ \sum_{l=1}^{k} \sum_{u \in U_{l}} \Delta^{+}_{uw}(x_{l}-y)
N(W_{U_{1}}(x_{1}),\dots,W_{U_{l}-\{u\}},\dots,W_{U_{k}}(x_{k}));
\eea
\bea
<\Omega, N(W_{U_{1}}(x_{1}),\dots,W_{U_{k}}(x_{k}))
N(W_{U_{k+1}}(x_{k+1}),\dots,W_{U_{n}}(x_{n}),\varphi_{w}(y)) \Omega>
\nonumber \\
= \sum_{l=1}^{k} \sum_{u \in U_{l}} 
<\Omega,\varphi_{u}(x_{l}) \varphi_{w}(y) \Omega> \times
\nonumber \\
<\Omega, N(W_{U_{1}}(x_{1}),\dots,W_{U_{l}-\{u\}}(x_{l}),
\dots,W_{U_{k}}(x_{k}))
N(W_{U_{k+1}}(x_{k+1}),\dots,W_{U_{n}}(x_{n})) \Omega>.
\eea
\label{normal2}
\end{cor}
{\bf Proof:}
We take in Corollary \ref{normal1}
$\{u_{1},\dots,u_{n}\} = \cup_{i=1}^{k} U_{i}$
and we ``colapse" the variables
$x_{j}$
pertaining to the same set 
$U_{i}$. 
In this way the first two relations follow.  The last relation follows from the
first one.
$\qed$

The relation (\ref{normal-type}) from the proof of Proposition
\ref{multi-normal} is remarkable and deserves a special name. We call an
ensemble of operator-valued distributions
$E(W_{U_{1}}(x_{1}),\dots,W_{U_{n}}(x_{n}))$
{\it of (supersymmetric) Wick type} if and only if the following two
conditions are verified:
\bea
E(\emptyset,\dots,\emptyset)) = {\rm const.}, \quad
\nonumber \\
E(\emptyset,\dots,\varphi_{u}(x_{l}),\dots,\emptyset) = 
<\Omega, E(\emptyset,\dots,\varphi_{u}(x_{l}),\dots,\emptyset)\Omega > {\bf 1}
+ E(\emptyset,\dots,\emptyset) \varphi_{u}(x_{l}),
\eea
\bea
[ E(W_{U_{1}}(x_{1}),\dots,W_{U_{n}}(x_{n})) , \varphi_{w}(y) ]
\nonumber \\
= \sum_{l=1}^{n} \sum_{u \in U_{l}}
E(W_{U_{1}}(x_{1}),\dots,W_{U_{l}-\{u\}},\dots,W_{U_{n}}(x_{n})) 
\Delta_{uw}(x_{l}-y);
\label{wick-type}
\eea

It is easy to note that if
$E(W_{U_{1}}(x_{1}),\dots,W_{U_{n}}(x_{n}))$
and
$F(W_{U_{n+1}}(x_{n+1}),\dots,W_{U_{n+m}}(x_{n+m}))$
are expression of Wick type, then
$
E(W_{U_{1}}(x_{1}),\dots,W_{U_{n}}(x_{n}))
F(W_{U_{n+1}}(x_{n+1}),\dots,W_{U_{n+m}}(x_{n+m}))
$
is also an expression of Wick type. The assertion stays true for more than two
factors.

We can extend by linearity a expression of Wick type to Wick polynomials: if
$W_{j}(x_{j})$
are Wick monomials, and
$c_{i_{1}j_{1}}, \quad c_{i_{n}j_{n}} \in {\cal G}$
then we define:
\bea
E(\sum c_{i_{1}j_{1}} W_{j_{1}}(x_{1}),\dots,
\sum c_{i_{n}j_{n}} W_{j_{n}}(x_{n}))
\equiv \sum c_{i_{1}j_{1}} \cdots c_{i_{n}j_{n}}
E(W_{j_{1}}(x_{1}),\dots,W_{j_{n}}(x_{n})) \quad
\eea
where the convention about the order of factors is important because of the
non-commutativity of the elements of the Grassmann algebra.

The well-known $0$-theorem of Epstein-Glaser asserts that expressions of the
type
\be
E(W_{1}(x_{1}),\dots,W_{n}(x_{n})) \equiv d(x_{1},\dots,x_{n})
N(W_{1}(x_{1}),\dots,W_{n}(x_{n}))
\ee
where
$
d(x_{1},\dots,x_{n})
$
is a translation-invariant distribution, are well-defined. They are obviously
expressions of Wick type. If we take the distribution $d$ of the form
\be
d(x_{1},\dots,x_{n}) = p(\partial) \delta^{n-1}(X)
\label{ano}
\ee
where
$p(\partial)$
is a polynomial in the partial derivatives and
\be
\delta^{n-1}(X) \equiv \delta(x_{1}-x_{n}) \cdots \delta(x_{n-1}-x_{n})
\ee
then we obtain some special expressions of Wick type called {\it quasi-local
operators} \cite{BS2}. Such type of operator have a distinguished r\^ole in
perturbative renormalization theory.

This analysis culminates with an extremely neat form of Wick theorem.
\begin{thm}
Let
$E(W_{U_{1}}(x_{1}),\dots,W_{U_{n}}(x_{n}))$
be an expression of Wick type. Then the following relation is valid:
\bea
E(W_{U_{1}}(x_{1}),\dots,W_{U_{n}}(x_{n}))
\nonumber \\
= \sum_{U^{\prime}_{i} \subset U_{i}}
<\Omega, E(W_{CU^{\prime}_{1}}(x_{1}),\dots,W_{CU^{\prime}_{n}}(x_{n}))\Omega>
N(W_{U^{\prime}_{1}}(x_{1}),\dots,W_{U^{\prime}_{n}}(x_{n}));
\label{wick1}
\eea
here
$CU^{\prime}_{i} \equiv U_{i} - U^{\prime}_{i}$
are the set-theoretically complements.
\label{wick-thm}
\end{thm}
{\bf Proof:}

It is done by induction over the rank
$r \equiv |U_{1}| + \cdots |U_{n}|$.
For $r = 1$ the formula from the statement is trivial. We suppose that the
formula is true for
$|U_{1}| + \cdots |U_{n}| = r-1$
and we prove it for
$|U_{1}| + \cdots |U_{n}| = r.$
One commutes both sides of the identity to be proven with an arbitrary
$\varphi_{w}(y)$
and, using the induction hypothesis, obtains equality. It follows that the
relation to be proven is valid up to a constant operator. If we average on the
vacuum we obtain that the constant is, in fact, zero.
$\qed$

\subsection{Derivatives of Wick Polynomials}

We can give alternative expressions for this theorem if we introduce the notion
of derivative of a Wick monomial \cite{DF1}. We give here a more compact
treatment. Let us denote the coordinates on the dual of the classical fields
bundle
$(J^{r}(\R^{N},M))^{*}$
by
$v^{A}_{\alpha}$;
the duality form is:
\be
<v,u> \equiv \sum_{A,\alpha} v^{A}_{\alpha} u_{A}^{\alpha}.
\ee

We consider the polar of the mass-shell:
\be
{\cal M}^{0} \equiv \{ v \in (J^{r}(\R^{N},M))^{*} \vert <v,u> = 0, \quad
\forall u \in J^{r}(\R^{N},M) \}.
\ee

Then we have the following elementary result:
\begin{prop}
Let
$ v \in {\cal M}^{0}$;
the the operator
$\rho(v): {\rm sWick}({\cal F}) \rightarrow {\rm sWick}({\cal F})$
is well defined by
\be
\rho(v) W_{U}(x) \equiv \sum_{u \in U} \quad <v,u> \quad W_{U - \{u\}}(x)
\ee
and linearity. Moreover, these operators commute among themselves:
\be
[ \rho(v_{1}), \rho(v_{2}) ] = 0, \quad \forall v_{1}, v_{2} \in {\cal M}^{0}.
\ee
\end{prop}

We call
$\rho(v)$
{\it derivative operators} of Wick polynomials. Because of the commutativity it
makes sense to define for any set 
$V = \{v_{1},\dots,v_{n}\}$
of elements from
${\cal M}^{0}$
derivative operators of higher order through:
\be
\rho(V) \equiv \prod_{i=1}^{n} \rho(v_{i}).
\ee

We can provide an alternative expression for the normal products and Wick
monomials. 
\begin{prop}
Let us consider
$\{v_{j}\}_{j \in J}$
a basis in
${\cal M}^{0}$
and
$\{v^{*}_{j}\}_{j \in J}$
a dual basis in a supplement
${\cal M}^{\prime}$
of
${\cal M} \subset J^{r}(\R^{N},M)$
such that the completeness relation is valid:
\be
\sum_{j \in J} (v^{*}_{j})^{\alpha}_{A} (v_{j})_{\beta}^{B} 
= \delta^{\alpha}_{\beta} \delta^{B}_{A}.
\ee
Then the following formul\ae~ are valid:
\be
N(\varphi_{u_{1}}(x_{1}),\dots,\varphi_{u_{n}}(x_{n}))
= \sum_{j_{1},\dots,j_{n} \in J} \prod_{k=1}^{n}<v_{j_{k}}, u_{k}>
N(\varphi_{v^{*}_{j_{1}}}(x_{1}),\dots,\varphi_{v^{*}_{j_{n}}}(x_{n}))
\ee
\be
W_{u_{1},\dots,u_{n}}(x) = \sum_{j_{1},\dots,j_{n} \in J}
\prod_{k=1}^{n} <v_{j_{k}},u_{k}> W_{v_{j_{1}}^{*},\dots,v^{*}_{j_{n}}}(x).
\ee
\label{normal3}
\end{prop}
{\bf Proof:}
We use a technique familiar by now. Let us denote the right hand side of the 
first relation by
$N^{\prime}(\varphi_{u_{1}}(x_{1}),\dots,\varphi_{u_{n}}(x_{n}))$
and check that the properties from Proposition \ref{normal} are true. One must
use the relation
\be
\sum_{j \in J} <v_{j}, u> \Delta_{v^{*}_{j}w}(x-y) = \Delta_{uw}(x-y)
\label{complete}
\ee
which is a consequence of the completeness relation. The second relation from
the statement follows if we ``colapse" the arguments into the first one.
$\qed$

Now we can give two alternative formulation of Wick theorem. First we have:
\begin{thm}
Every Wick expression
$E(W_{1}(x_{1}),\dots,W_{n}(x_{n}))$
(here
$W_{1}(x_{1}),\dots,W_{n}(x_{n})$
are Wick polynomials) verify the following relation:
\bea
[ E(W_{1}(x_{1}),\dots,W_{n}(x_{n})), \varphi_{w}(y) ]
\nonumber \\
= \sum_{l=1}^{n} \sum_{j\in J} \Delta_{v^{*}_{j}w}(x_{l}-y) \quad
E(W_{1}(x_{1}),\dots,\rho(v_{j}) W_{l}(x_{l}),\dots,W_{n}(x_{n})).
\eea

In particular we have for every Wick polynomial:
\be
[ W(x), \varphi_{w}(y) ]
= \sum_{j\in J} \Delta_{v^{*}_{j}w}(x-y) \rho(v_{j}) W(x).
\ee
\label{wick-der}
\end{thm}
{\bf Proof:}
It is sufficient to consider that
$W_{1}(x_{1}),\dots,W_{n}(x_{n})$
are Wick monomials. Then we use the defining relation for a expression of Wick
type and the relation (\ref{complete}).
$\qed$

Now we can give another compact form of Wick theorem.
\begin{thm}
The following formula is valid:
\bea
E(W_{1}(x_{1}),\dots,W_{n}(x_{n}))
\nonumber \\
= \sum_{V_{i}} 
<\Omega,E(\rho(V_{1}) W_{1}(x_{1}),\dots,\rho(V_{n}) W_{n}(x_{n})) \Omega> 
\quad
N(W_{V_{1}^{*}}(x_{1}),\dots,W_{V_{n}^{*}}(x_{n}));
\label{wick2}
\eea
where the sum runs over all sets 
$V_{i}$
of elements of the type
$v_{j} \quad (j \in J)$
from 
${\cal M}^{0}$.
\end{thm}
{\bf Proof:}
As before, it is sufficient to consider that the expressions
$W_{i}$
are Wick monomials. If we use Proposition \ref{normal3} we obtain that the
right hand side of the relation from the statement coincides with the right
hand side of the relation from Wick theorem (\ref{wick1}).
$\qed$

We can extend the operation of derivation
$\partial_{\alpha}$
to elements of the polar
${\cal M}^{0}$
through duality: we have
\begin{prop}
Let us define
$
\partial_{\alpha}: (J^{r}(\R^{N}, M))^{*} \rightarrow 
(J^{r-|\alpha|}(\R^{N}, M))^{*}
$
according to:
\be
(\partial_{\alpha}v)_{\beta}^{A} \equiv v^{\alpha+\beta}_{A}.
\ee
Then: 
(i) If 
$v \in {\cal M}^{0}$ 
we also have
$\partial_{\alpha}v \in {\cal M}^{0}$
for any multi-index $\alpha$.

(ii) The following commutation relation is valid:
\be
[ \rho(v) , \partial_{\alpha} ] = \rho(\partial_{\alpha}v).
\label{partial-a}
\ee
\end{prop}
{\bf Proof:}
The first assertion follows from the corresponding property for the derivatives
of the elements of
$J^{r}(\R^{N}, M)$
and the duality relation:
\be
<\partial_{\alpha}v,u> = <v, \partial_{\alpha}u>.
\ee
This identity, as well as the last relation from the statement can be proved
directly from the definitions by elementary computations.
$\qed$

It it convenient to introduce some particular derivatives of the type
$\rho(v)$.
We consider some field
$\varphi^{A}$
(the index $A$ is fixed) constrained only by the Klein-Gordon equation
(\ref{KG}). Then we define the elements
$v_{A}, \quad v_{A}^{\mu} \in {\cal M}^{0}$
by giving only the {\bf non-zero} entries:
\be
(v_{A})^{B} = \delta^{B}_{A} \quad,
(v_{A})^{B}_{\nu\rho} 
= - {1\over 4} M_{A}^{2} \delta^{B}_{A} g_{\nu\rho}
\ee
and respectively
\be
(v_{A}^{\mu})^{B}_{\nu} = \delta^{B}_{A} \delta^{\mu}_{\nu}, \quad
(v_{A}^{\mu})^{B}_{\nu\rho\sigma} = - {1\over 4} M_{A}^{2}
\delta^{B}_{A} {\cal S}_{\nu\rho\sigma} \delta^{\mu}_{\nu} g_{\rho\sigma};
\ee
indeed we immediately have
$$
< v_{A}, u > = 0, \quad < v_{A}^{\mu}, u > = 0, \quad \forall u \in {\cal M}.
$$

Then we denote:
\be
{\partial \over \partial \varphi^{A}} \equiv \rho(v_{A}), \quad
{\partial \over \partial \varphi^{A}_{\mu}} \equiv \rho(v_{A}^{\mu});
\label{der-A}
\ee
these are in fact derivatives with respect to the basic fields and their first
order jet extension. 

Next we have the following result \cite{DF1} following directly from the second
formula of theorem \ref{wick-der}. We say that
$W_{U}$
is a Wick monomial of {\it first order} if all elements
$u \in U$
verify
$u_{A}^{\alpha} = 0, \forall |\alpha| > 1$.
A {\it first order} Wick polynomial is a sum of first order Wick monomials.
\begin{prop}
Suppose that $W$ is a Wick polynomial of first order. Then we have:
\be
[ W(x), \varphi^{A}(y) ]
= {\partial \over \partial \varphi^{B}} W(x) \Delta^{AB}(x-y) \rho(v_{j}) 
+ {\partial \over \partial \varphi^{B}_{\mu}} W(x) 
\partial_{\mu}\Delta^{AB}(x-y).
\ee
\end{prop}
The proof is elementary. We also define:
\be
\partial_{\sigma}\cdot {\partial \over \partial \varphi^{A}} \equiv 
\rho(\partial_{\sigma}\cdot v_{A}), \quad
\partial_{\sigma}\cdot{\partial \over \partial \varphi^{A}_{\mu}} \equiv 
\rho(\partial_{\sigma}\cdot v_{A}^{\mu}).
\label{der}
\ee

Then we have for any first order Wick polynomial $W$ the following formul\ae:
\be
\partial_{\sigma}\cdot {\partial \over \partial \varphi^{A}} W(x)
= - {1\over 4} M_{A}^{2} {\partial \over \partial \varphi^{A}_{\sigma}} W(x),
\quad
\partial_{\sigma}\cdot {\partial \over \partial \varphi^{A}_{\mu}} W(x)
= \delta^{\mu}_{\sigma} {\partial \over \partial \varphi^{A}} W(x).
\label{der1}
\ee

We finally notice that we can define in a natural way the ghost number of
the derivatives
$
{\partial \over \partial \varphi^{A}}, \quad
{\partial \over \partial \varphi^{A}_{\mu}}
$
to be
$z_{A}$.
If the elements of the set
$V = \{v_{1},\dots,v_{k}\}$
are of defined ghost number, then
\be
gh(V) \equiv \sum gh(v_{i}).
\ee

\subsection{Wick Monomials\label{wick}}

In this Subsection, we make the connection with the ordinary Wick monomials
defined in the original Fock space ${\cal F}$. Loosely speaking, if we strip a
supersymmetric Wick monomial of their Grassmann factors in a consistent way, we
obtain the usual Wick monomials. First we have:
\begin{prop}
Let
$\sigma$
be section of the fibre bundle:
$
J^{r}(\R^{N}, M) \rightarrow J^{r}(\R^{N}, M)/{\cal M}.
$
Then every Wick monomial can be uniquely written in the form
\be
W_{u_{1},\dots,u_{n}}(x) 
= \prod_{i=1}^{n} (\widetilde{u_{i}})^{\alpha_{i}}_{A_{i}} \quad
W^{A_{1},\dots,A_{n}}_{\alpha_{1},\dots,\alpha_{n}}(x)
= \prod_{i=1}^{n} (\widetilde{u_{i}})^{\alpha_{i}}_{A_{i}} g_{A_{i}} \quad
{\cal W}^{A_{1},\dots,A_{n}}_{\alpha_{1},\dots,\alpha_{n}}(x)
\ee
where
$
\widetilde{u_{i}} \equiv \sigma([u_{i}]),
$
$
W^{A_{1},\dots,A_{n}}_{\alpha_{1},\dots,\alpha_{n}}(x)
$
are operator-valued distributions with values in
$
{\cal G} \otimes {\cal L}({\cal F})
$
and
$
{\cal W}^{A_{1},\dots,A_{n}}_{\alpha_{1},\dots,\alpha_{n}}(x)
$
are operator-valued distributions with values in
$
{\cal L}({\cal F}).
$
A similar writing can be established for supersymmetric expressions of Wick
type.
\end{prop}
{\bf Proof:}
We define recurrently the expressions
$
W^{A_{1},\dots,A_{n}}_{\alpha_{1},\dots,\alpha_{n}}(x)
$
through the following properties:
\be
W^{A_{1},\dots,A_{n}}_{\alpha_{1},\dots,\alpha_{n}}(x) \Omega
= \prod_{i=1}^{n} \partial_{\alpha_{i}} \varphi^{A_{i}}(x) \Omega;
\ee
\be
[ W^{A_{1},\dots,A_{n}}_{\alpha_{1},\dots,\alpha_{n}}(x), 
\partial_{\beta}\varphi^{B}(x) ]
= \sum_{l=1}^{n} W^{A_{1},\dots,\widehat{A_{l}},\dots,A_{n}}_{\alpha_{1},
\dots,\widehat{\alpha_{l}},\dots,\alpha_{n}}(x)
\partial_{\alpha_{l}}^{x} \partial_{\beta}^{y} \Delta^{A_{l}B}(x-y);
\ee
\be
W^{\emptyset}_{\emptyset}(x) = {\bf 1}.
\ee

One can prove that these relations define uniquely the expressions
$
W^{A_{1},\dots,A_{n}}_{\alpha_{1},\dots,\alpha_{n}}(x)
$
using a familiar argument. Then we obtain the first equality from the statement
using the uniqueness argument from Proposition \ref{wick-def}.  

The expressions
$
{\cal W}^{A_{1},\dots,A_{n}}_{\alpha_{1},\dots,\alpha_{n}}(x)
$
can be defined quite similarly and we obtain the second equality from the
statement.
$\qed$

One can see that the expressions 
$
W^{A_{1},\dots,A_{n}}_{\alpha_{1},\dots,\alpha_{n}}(x)
$
are in fact supersymmetric Wick monomials: they can be obtained for some
special choice of the classical fields
$u_{i}$.
Moreover, they are completely symmetric in the couples
$(A,\alpha)$
and are not linearly independent. In fact, we have ``equation of motion" of the
type:
\be
\sum_{\alpha_{l}} u^{\alpha_{l}}_{A_{l}}
W^{A_{1},\dots,A_{n}}_{\alpha_{1},\dots,\alpha_{n}}(x) = 0, \quad
\forall u \in {\cal M}.
\ee

Similar assertion are valid for
$
{\cal W}^{A_{1},\dots,A_{n}}_{\alpha_{1},\dots,\alpha_{n}}(x);
$
more precisely, we have skew-symmetry in the couples
$(A,\alpha)$
and appropriate equations of motion. The expression
$
{\cal W}^{A_{1},\dots,A_{n}}_{\alpha_{1},\dots,\alpha_{n}}(x)
$
are called {\it Wick monomials} and
$
W_{u_{1},\dots,u_{n}}(x)
$
is the {\it associated supersymmetric Wick monomial}. A {\it Wick polynomial}
is a operator acting in ${\cal F}$ of the following form:
\be
{\cal L}(x) = \sum 
{\cal C}_{A_{1},\dots,A_{n}}^{\alpha_{1},\dots,\alpha_{n}}
{\cal W}^{A_{1},\dots,A_{n}}_{\alpha_{1},\dots,\alpha_{n}}(x)
\ee
where 
$
{\cal C}_{A_{1},\dots,A_{n}}^{\alpha_{1},\dots,\alpha_{n}}
$
are complex constants with convenient (anti)-symmetry properties. We denote by
Wick(${\cal F}$)
the set of Wick polynomials in ${\cal F}$. If we express the operators
$
{\cal W}^{A_{1},\dots,A_{n}}_{\alpha_{1},\dots,\alpha_{n}}(x)
$
in term of
$
W^{A_{1},\dots,A_{n}}_{\alpha_{1},\dots,\alpha_{n}}(x)
$
then we canonically associate to the Wick polynomial
${\cal L}(x)$
a super-symmetric Wick polynomial
$L(x)$
acting in
${\cal G} \otimes {\cal L}({\cal F}).$

We can define now some derivative operators. First, we note that the derivation
$\rho(v)$
induces a derivation, also denoted  
$\rho(v)$
on the space of Wick polynomials. Next, we have the following result:
\begin{prop}
Let us define the operators:
$\partial^{\beta}_{B}: {\rm sWick} \rightarrow {\rm sWick}$
according to:
\be
\partial^{\beta}_{B} W^{A_{1},\dots,A_{n}}_{\alpha_{1},\dots,\alpha_{n}}(x)
\equiv \sum_{l=1}^{n} \delta^{A_{l}}_{B} \delta^{\beta}_{\alpha_{l}}
W^{A_{1},\dots,\widehat{A_{l}},\dots,A_{n}}_{\alpha_{1},\dots,
\widehat{\alpha_{l}},\dots,\alpha_{n}}(x).
\ee
Then the following relations are true:
\be
\partial^{\beta}_{B} W_{U}(x) = \sum_{u \in U} \tilde{u}^{\beta}_{B}
W_{U-\{u\}}(x), \quad
\rho(v) = \sum_{A,\alpha} v^{A}_{\alpha} \partial^{\alpha}_{A}, \quad
\forall v \in {\cal M}^{0}.
\ee
\end{prop}

\newpage

\section{Perturbation Theory in the Causal Approach\label{pert}}
 
We give here the essential ingredients of perturbation theory using the
supersymmetric formalism described in the preceding Section. 

\subsection{Bogoliubov Axioms}{\label{bogoliubov}}

We use, essentially, the point of view of of Stora and Fredenhagen \cite{Sto1},
\cite{DF1}, \cite{Bo} using chronological product. An equivalent point of view
uses retarded products \cite{St1}. By a {\it perturbation theory} in the sense
of Bogoliubov we mean an ensemble of operator-valued distributions
$
T(W_{1}(x_{1}),\dots,W_{n}(x_{n})) \in {\cal G} \otimes {\cal L}({\cal F}),
\quad n = 1,2,\dots
$
called {\it (super-symmetric) chronological products} (here
$
W_{1}(x_{1}),\dots,W_{n}(x_{n})
$
are supersymmetric Wick polynomials) verifying the following set of axioms:
\begin{itemize}
\item
Symmetry in all arguments
$
W_{1}(x_{1}),\dots,W_{n}(x_{n});
$
\item
Poincar\'e invariance: for all 
$(a,L) \in inSL(2,\C)$
we have:
\be
U_{a, L} T(W_{1}(x_{1}),\dots,W_{n}(x_{n})) U^{-1}_{a, L} =
T(L\cdot W_{1}(\delta(L)\cdot x_{1}+a),\dots,
L\cdot W_{n}(\delta(L)\cdot x_{n}+a));
\label{invariance}
\ee

Sometimes it is possible to supplement this axiom by corresponding invariance
properties with respect to inversions (spatial and temporal) and charge
conjugation. For the standard model only the PCT invariance is available.
Also some other global symmetry with respect to some internal symmetry group
might be imposed.
\item
Causality: if
$x_{i} \geq x_{j}, \quad \forall i \leq k, \quad j \geq k+1$
then we have:
\be
T(W_{1}(x_{1}),\dots,W_{n}(x_{n})) =
T(W_{1}(x_{1}),\dots,W_{k}(x_{k})) T(W_{k+1}(x_{k+1}),\dots,W_{n}(x_{n}));
\label{causality}
\ee
\item
Unitarity: We define the {\it (super-symmetric) anti-chronological products}
according to
\be
(-1)^{n} \bar{T}(W_{1}(x_{1}),\dots,W_{n}(x_{n})) \equiv \sum_{r=1}^{n} 
(-1)^{r} \sum_{I_{1},\dots,I_{r} \in Part(\{1,\dots,n\})}
T_{I_{1}}(X_{1})\cdots T_{I_{r}}(X_{r})
\label{antichrono}
\ee
where the we have used the notation:
\be
T_{\{i_{1},\dots,i_{k}\}}(x_{i_{1}},\dots,x_{i_{k}}) \equiv 
T(W_{i_{1}}(x_{i_{1}}),\dots,W_{i_{k}}(x_{i_{k}})).
\ee
Then the unitarity axiom is:
\be
\bar{T}(W_{1}(x_{1}),\dots,W_{n}(x_{n}))
= T(W^{*}_{1}(x_{1}),\dots,W^{*}_{n}(x_{n}))
\label{unitarity}
\ee
\item
The ``initial condition"
\be
T(W(x)) = W(x).
\ee
\end{itemize}
\begin{rem}
From (\ref{causality}) one can derive easily that if we have
$x_{i} \sim x_{j}, \quad \forall i \leq k, \quad j \geq k+1$
then:
\be
[ T(W_{1}(x_{1}),\dots,W_{k}(x_{k})), T(W_{k+1}(x_{k+1}),\dots,W_{n}(x_{n})) ] 
= 0.
\label{commute}
\ee
\end{rem}

\newpage
\subsection{Epstein-Glaser Construction\label{eg}}

Epstein-Glaser construction provides an explicit solution for Bogoliubov
axioms. We sketch briefly the proof.
\begin{thm}
There exists a solution of Bogoliubov axioms.
\end{thm}
{\bf Proof:}
Goes by induction. One suppose that the chronological products are constructed 
up to the order $n-1$ such that all Bogoliubov axioms are verified. We
supplement the induction hypothesis with the requirement that the chronological
products (up to the order $n-1$) are expressions of Wick type; this means that
we have for all
$p = 1,\dots,n-1$:
\bea
[ T(W_{U_{1}}(x_{1}),\dots,W_{U_{p}}(x_{p})) , \varphi_{w}(y) ]
\nonumber \\
= \sum_{l=1}^{p} \sum_{u \in U_{l}}
T(W_{U_{1}}(x_{1}),\dots,W_{U_{l}-\{u\}},\dots,W_{U_{p}}(x_{p})) 
\Delta_{uw}(x_{l}-y)
\label{wick-chrono1}
\eea
so, according to Wick theorem \ref{wick-thm} we have the expansion:
\bea
T(W_{U_{1}}(x_{1}),\dots,W_{U_{p}}(x_{p}))
\nonumber \\
= \sum_{U^{\prime}_{i} \subset U_{i}}
<\Omega, T(W_{CU^{\prime}_{1}}(x_{1}),\dots,W_{CU^{\prime}_{p}}(x_{p}))\Omega>
N(W_{U^{\prime}_{1}}(x_{1}),\dots,W_{U^{\prime}_{p}}(x_{p})).
\label{wick-chrono2}
\eea

We can also include in the induction hypothesis a limitation on the order of
singularity of the vacuum averages of the chronological products associated to
arbitrary Wick monomials
$W_{1},\dots,W_{p}$;
explicitly:
\be
\omega(<\Omega, T(W_{1}(x_{1}),\dots,W_{p}(x_{p}))\Omega>) \leq
\sum_{l=1}^{p} \omega(W_{l}) - 4(p-1), \quad p = 1,\dots,n-1
\label{power}
\ee
where by
$\omega(d)$
we mean the order of singularity of the (numerical) distribution $d$ and by
$\omega(W)$
we mean the canonical dimension of the Wick monomial $W$. It is easy to check
that the induction hypothesis is true for 
$n = 1$.

The construction of Epstein-Glaser is based on the commutator
$D(W_{U_{1}}(x_{1}),\dots,W_{U_{n}}(x_{n}))$
with causal support. The explicit expression of this commutator is known
in terms of the chronological products up to the order $n-1$. Moreover, from
the explicit formula it is clear that this expression is also of Wick type (it
is a sum of products of expressions of Wick type -according to the induction
hypothesis). So, Wick theorem can be applied and gives an expression of the
type: 
\bea
D(W_{U_{1}}(x_{1}),\dots,W_{U_{n}}(x_{n}))
\nonumber \\
= \sum_{U^{\prime}_{i} \subset U_{i}}
<\Omega, D(W_{CU^{\prime}_{1}}(x_{1}),\dots,W_{CU^{\prime}_{n}}(x_{n}))\Omega>
N(W_{U^{\prime}_{1}}(x_{1}),\dots,W_{U^{\prime}_{n}}(x_{n})).
\label{wick-d}
\eea

One can show that the order of singularity of the numerical distributions
\be
d(x_{1},\dots,x_{n})
\equiv <\Omega, D(W_{U_{1}}(x_{1}),\dots,W_{U_{n}}(x_{n}))\Omega>
\ee
verifies a restriction of the type (\ref{power}); this is the content of the
so-called {\it power-counting theorem.} Next, one can provide in a standard way
a causal splitting of the distribution $d(x_{1},\dots,x_{n})$ such that
Poincar\'e covariance and the order of singularity are preserved. This induces
a causal splitting for the operator-valued distribution
\be
D(W_{U_{1}}(x_{1}),\dots,W_{U_{n}}(x_{n})) 
= A(W_{U_{1}}(x_{1}),\dots,W_{U_{n}}(x_{n})) 
- R(W_{U_{1}}(x_{1}),\dots,W_{U_{n}}(x_{n}))
\ee
which can be used to construct the $n$-order chronological product
$T(W_{U_{1}}(x_{1}),\dots,W_{U_{n}}(x_{n}))$
The unitarity can be also fixed quite elementary \cite{EG1}. The induction is
finished.
$\qed$

From the construction it follows that one can define the chronological products
such that we have (\ref{wick-chrono1}), (\ref{wick-chrono2}) and (\ref{power})
for all
$p = 1,2,\dots$
The first relation is the normalisation condition ({\bf N3}) of \cite{DF1},
\cite{Bo}. According to the previous Section, we also have alternative
formulations for the first two of them, namely: we have for all 
$n \in \N$:
\bea
[ T(W_{1}(x_{1}),\dots,W_{n}(x_{n})), \varphi_{w}(y) ]
\nonumber \\
= \sum_{l=1}^{n} \sum_{j\in J} \Delta_{v^{*}_{j}w}(x_{l}-y) \quad
T(W_{1}(x_{1}),\dots,\rho(v_{j}) W_{l}(x_{l}),\dots,W_{n}(x_{n}))
\eea
and
\bea
T(W_{1}(x_{1}),\dots,W_{n}(x_{n}))
\nonumber \\
= \sum_{V_{i} \subset {\cal M}^{0}} <\Omega,
T(\rho(V_{1})W_{1}(x_{1}),\dots,\rho(V_{n})W_{n}(x_{n})) \Omega> \quad
N(W_{V_{1}^{*}}(x_{1}),\dots,W_{V_{n}^{*}}(x_{n})).
\label{wick-chrono3}
\eea

We still have some freedom on the chronological products which can be used to
impose another condition. Let
\be
\Delta_{uw} = \Delta^{adv}_{uw} - \Delta^{ret}_{uw}
\ee
be a causal splitting of the distribution with causal support
$\Delta_{uw}$.
By definition the {\it Feynman propagator} and the {\it Feynman antipropagator}
are:
\be
\Delta^{F}_{uw} \equiv 
\Delta^{adv}_{uw} - \Delta^{-}_{uw} = \Delta^{ret}_{uw} + \Delta^{+}_{uw},
\quad
\Delta^{AF} \equiv 
\Delta^{+}_{uw} - \Delta^{adv}_{uw} = - \Delta^{ret}_{uw} - \Delta^{-}_{uw}.
\label{propagator}
\ee

Then we have the following result \cite{Sto1}:
\begin{thm}
Suppose that the chronological products have been chosen such that they are
expressions of Wick type. Then they can be chosen such that one also has for
all
$n \in \N$:
\bea
T(W_{U_{1}}(x_{1}),\dots,W_{U_{n}}(x_{n}),\varphi_{w}(y))
\nonumber \\
= \sum_{U^{\prime}_{i} \subset U_{i}}
<\Omega, T(W_{CU^{\prime}_{1}}(x_{1}),\dots,W_{CU^{\prime}_{n}}(x_{n}))\Omega>
N(W_{U^{\prime}_{1}}(x_{1}),\dots,W_{U^{\prime}_{n}}(x_{n}),\varphi_{w}(y))
\nonumber \\
+ \sum_{l=1}^{n} \sum_{u \in U_{l}} \Delta^{F}_{uw}(x_{l}-y) \quad
T(W_{U_{1}}(x_{1}),\dots,W_{U_{l}-\{u\}},\dots,W_{U_{n}}(x_{n}))
\label{wick-chrono4}
\eea
\end{thm}
{\bf Proof:}
It is also based on induction on the rank
$|U_{1}| + \cdots + |U_{n}|$
(see also \cite{Bo}). One can easily see that the relation from the
statement is trivial for 
$r = 1$.
We suppose that it is true for
$|U_{1}| + \cdots |U_{n}| = r-1$
and prove that it can be fixed for 
$|U_{1}| + \cdots |U_{n}| = r$
also. We use a familiar technique, namely we consider for this case the
commutator of both sides of (\ref{wick-chrono3}) with an arbitrary
$\varphi_{w'}(z)$
and, using the induction hypothesis, we get zero. So, the relation from the
statement for
$|U_{1}| + \cdots |U_{n}| = r$
can be affected by a constant ``anomaly":
\bea
c(x_{1},\dots,x_{n},y)
\equiv 
<\Omega, T(W_{U_{1}}(x_{1}),\dots,W_{U_{n}}(x_{n}),\varphi_{w}(y))\Omega>
\nonumber \\
- \sum_{l=1}^{n} \sum_{u \in U_{l}} \Delta^{F}_{uw}(x_{l}-y) \quad
<\Omega,T(W_{U_{1}}(x_{1}),\dots,W_{U_{l}-\{u\}},\dots,W_{U_{n}}(x_{n}))
\Omega>.
\label{c}
\eea

Using the causal factorisation property of the chronological products, the
induction hypothesis and property (\ref{wick-chrono2}) one can prove that 
the support of the distribution
$c(x_{1},\dots,x_{n},y)$
is contained in the diagonal set
$x_{1}= \dots =x_{n} = y$.
This means that we have the generic form
\be
c(x_{1},\dots,x_{n},y) = p(\partial) \delta(x_{1}-y) \cdots \delta(x_{n}-y)
\ee
with
$p(\partial)$
some polynomials in the partial derivatives. Moreover, this numerical
distribution has convenient covariance properties and a limitation on the
degree of $p$ comes from the power counting limitations in the right hand side
of (\ref{c}). In the end, it follows that we can absorb the anomaly
$c(x_{1},\dots,x_{n},y)$
into the vacuum sector of the chronological product
$
T(W_{U_{1}}(x_{1}),\dots,W_{U_{n}}(x_{n}),\varphi_{w}(y)), \quad
|U_{1}| + \cdots |U_{n}| = r
$
without affecting the Epstein-Glaser induction construction.
$\qed$

The relation appearing in this proposition is called in \cite{DF1}, \cite{Bo}
the normalization condition ({\bf N4}).

As in the preceding Section, one can define the chronological products acting
in the Hilbert space
${\cal F}$
by stripping the Grassmann variables.

\newpage
\section{General Gauge Theories}

\subsection{The Supersymmetric BRST Operator}

In the general setting of Subsection \ref{free} we define a {\it BRST operator}
$d_{Q}$
on the set of polynomials in the fields
$\phi^{A}_{\pm}$
through the following properties:
\begin{itemize}
\item
It gives zero on the constant operator:
\be
d_{Q} {\bf 1} = 0.
\ee
\item
It is linear over $\C$.
\item
It acts on the basic fields as follows:
\be
d_{Q} \phi^{A}_{\pm}(x) = - i \sum_{|\alpha| \leq s} \sum_{B} 
{(q^{\alpha})^{A}}_{B} \quad \partial_{\alpha} \phi^{B}_{\pm}(x)
\ee
where
$q^{\alpha}$
are real
$N \times N$
matrices constrained by
\be
{(q^{\alpha})^{A}}_{B} = 0, \quad iff \quad z_{B} - z_{A} \not= 1;
\ee
here $s \in \N^{*}$ is called the {\it rank} of the BRST operator. The usual 
case is
$s=1$.
\item
It is a (graded) derivative operator in the sense that for all
$\epsilon_{1},\dots,\epsilon_{n} = \pm$
we have:
\be
d_{Q} [ \phi^{A_{1}}_{\epsilon_{1}}(x_{1}) \dots 
\phi^{A_{n}}_{\epsilon_{n}}(x_{n}) ] = \sum_{l=1}^{n} 
\prod_{i <l} (-1)^{z_{A_{i}}} 
\phi^{A_{1}}_{\epsilon_{1}}(x_{1}) \dots 
d_{Q} \phi^{A_{l}}_{\epsilon_{l}}(x_{l})
\dots \phi^{A_{n}}_{\epsilon_{n}}(x_{n})
\ee
\item
It commutes with the derivative operators:
\be
[ d_{Q}, \partial_{\beta} ] = 0.
\ee
\end{itemize}

It is clear that the usual BRST operator appearing in Yang-Mills models is a
particular case of this general framework.  One can naturally extend the
operator $d_{Q}$ to the set of polynomials in the fields $\varphi^{A}_{\pm}$;
then the following properties are true:
\begin{itemize}
\item
It gives zero on the constant operator:
\be
d_{Q} {\bf 1} = 0.
\ee
\item
It is linear over ${\cal G}$.
\item
It acts on the basic fields as follows:
\be
d_{Q} \varphi^{A}_{\pm}(x) = - i \sum_{|\alpha| \leq s} \sum_{B} 
{(Q^{\alpha})^{A}}_{B} \quad \partial_{\alpha} \varphi^{B}_{\pm}(x)
\ee
where
\be
{(Q^{\alpha})^{A}}_{B} = g_{A} g_{B}^{-1} {(q^{\alpha})^{A}}_{B}
\ee
and
$deg({(Q^{\alpha})^{A}}_{B}) = - 1$.
\item
It is a derivative operator: 
\be
d_{Q} [ \varphi^{A_{1}}_{\epsilon_{1}}(x_{1}) \dots 
\varphi^{A_{n}}_{\epsilon_{n}}(x_{n}) ] = -i \sum_{l=1}^{n} 
{(Q^{\alpha})^{A_{l}}}_{B}
\varphi^{A_{1}}_{\epsilon_{1}}(x_{1}) \dots 
\partial_{\alpha}\varphi^{A_{l}}_{\epsilon_{l}}(x_{l})
\dots \varphi^{A_{n}}_{\epsilon_{n}}(x_{n})
\ee
\item
It commutes with the derivative operators:
\be
[ d_{Q}, \partial_{\beta} ] = 0.
\ee
\end{itemize}

We compute the action of the operator on the supersymmetric fields. We have:
\begin{prop}
The following formula is true:
\be
d_{Q} \varphi_{u}(x) = \zeta^{-1} \varphi_{\zeta Q\cdot u}(x)
\ee
where
$\zeta \in {\cal G}_{1}$
is a fixed invertible element and we have defined:
\be
(Q\cdot u)^{\alpha}_{A} \equiv - \sum_{\beta+\gamma=\alpha} \sum_{B}
{(Q^{\beta})^{B}}_{A} \quad u^{\gamma}_{B}.
\ee
\end{prop}

The proof is elementary. We have introduced the factor
$\zeta$
because
$(Q\cdot u)^{\alpha}_{A} \in {\cal G}_{-1}$
and in this way
$\zeta Q\cdot u$
has real values and it can be considered as an element of the classical
manifold
$J^{r}(\R^{N}, M).$

If we apply the operator
$d_{Q}$
to the last commutation relation (\ref{CCR-Grassmann}) we obtain:
\be
\Delta_{\zeta Q\cdot u,w} = - \Delta{u, \zeta Q\cdot w}.
\ee

Another consequence of the preceding proposition is:
\begin{cor}
The following formul\ae~ are true:
\be
d_{Q} [ \varphi_{u_{1}}^{\epsilon_{1}}(x_{1}) \dots 
\varphi_{u_{n}}^{\epsilon_{n}}(x_{n}) ] = i \zeta^{-1} \sum_{l=1}^{n} 
\varphi_{u_{1}}^{\epsilon_{1}}(x_{1}) \dots 
\varphi_{\zeta Q\cdot u_{l}}^{\epsilon_{l}}(x_{l})
\dots \varphi_{u_{n}}^{\epsilon_{n}}(x_{n}),
\ee
\be
d_{Q} N( \varphi_{u_{1}}^{\epsilon_{1}}(x_{1}) \dots 
\varphi_{u_{n}}^{\epsilon_{n}}(x_{n}) ) = i \zeta^{-1} \sum_{l=1}^{n} 
N( \varphi_{u_{1}}^{\epsilon_{1}}(x_{1}) \dots 
\varphi_{\zeta Q\cdot u_{l}}^{\epsilon_{l}}(x_{l})
\dots \varphi_{u_{n}}^{\epsilon_{n}}(x_{n}) ),
\ee
\be
d_{Q} W_{u_{1},\dots,u_{n}}(x) = i \zeta^{-1} \sum_{l=1}^{n}
W_{u_{1},\dots,\zeta Q\cdot u_{l},\dots,u_{n}}(x).
\ee
\be
d_{Q} W^{A_{1},\dots,A_{n}}_{\alpha_{1},\dots,\alpha_{n}}(x) =
- i \sum_{l=1}^{n} \sum_{B,\beta} {(Q^{\beta})^{A_{l}}}_{B}
W^{A_{1},\dots,A_{l-1},B,A_{l+1},\dots,A_{n}}_{\alpha_{1},\dots,
\alpha_{l-1},\beta+\alpha_{l},\alpha_{l+1},\dots,\alpha_{n}}(x)
\ee
\be
d_{Q} {\cal W}^{A_{1},\dots,A_{n}}_{\alpha_{1},\dots,\alpha_{n}}(x)
= - i \sum_{l=1}^{n} \sum_{B,\beta} \prod_{i < l} (-1)^{z_{A_{i}}}
{(q^{\beta})^{A_{l}}}_{B}
{\cal W}^{A_{1},\dots,A_{l-1},B,A_{l+1},\dots,A_{n}}_{\alpha_{1},\dots,
\alpha_{l-1},\beta+\alpha_{l},\alpha_{l+1},\dots,\alpha_{n}}(x).
\ee
\end{cor}
{\bf Proof:}
The first relation goes by direct computations from the derivative property of
the operator
$d_{Q}$,
the second relation follows from the first if we use Proposition
\ref{normal-pm} and the third one follows if we ``colapse" the variables into
the preceding one. The last two relations are direct consequences of the
definitions.
$\qed$

We will have to extend the action of the BRST operator to the dual space of
classical fields
$(J^{r}(\R^{N},M))^{*}$.
For this we need the following result:
\begin{prop}
(i) If
$u \in {\cal M}$
then
$\zeta Q\cdot u \in {\cal M}$.

(ii) If
$v \in (J^{r}(\R^{N},M))^{*}$
let us define
$Q\cdot v \in (J^{r-s}(\R^{N},M))^{*}$
according to
\be
(Q\cdot v)^{A}_{\alpha} \equiv \sum_{B,\beta} 
{(Q^{\beta})^{A}}_{B} v^{B}_{\alpha+\beta}.
\ee

Suppose now that
$v \in {\cal M}^{0}$;
then
$\zeta Q\cdot v \in {\cal M}^{0}$.

(iii) The following relation is valid:
\be
[ d_{Q}, \rho(v) ] = i \zeta^{-1} \rho(\zeta Q\cdot v)
\label{dQ-a}
\ee
\end{prop}
{\bf Proof:}
The proof of the first two assertions are based on elementary computations. For
the last relation, it is sufficient to prove it to be true when applied on a
Wick monomial
$W_{U}(x).$
$\qed$

\subsection{Gauge Invariant Models}

We generalize the framework outlined in the Introduction i.e. we suppose that
we have a set of Wick polynomials
${\cal A}^{i}(x), \quad i =1,\dots,p$,
which we organize as a Wick multiplet (a column matrix) 
${\cal A}$ 
and some
$p \times p$
matrices
$c^{\alpha}$
such that the following relation is true:
\be
d_{Q} {\cal A}(x) = i \sum_{\alpha} c^{\alpha} \partial_{\alpha} {\cal A}(x).
\ee

The we say that we have a {\it general gauge theory}. If
$A^{i}(x)$
are the supersymmetric Wick polynomials associated to
${\cal A}^{i}(x), \quad i =1,\dots,p$
then a similar relation is verified by them:
\be
d_{Q} A(x) = i \sum_{\alpha} c^{\alpha} \partial_{\alpha} A(x).
\ee
We have the following consequence:
\begin{prop}
Let
$A(x)$
be a general gauge theory. Then we also have:
\bea
d_{Q} [\rho(v_{1}) \cdots \rho(v_{k}) A(x)] = 
i \sum_{\alpha} c^{\alpha} \partial_{\alpha} 
[\rho(v_{1}) \cdots \rho(v_{k}) A(x)]
\nonumber \\
+ \sum_{l=1}^{k} \left[ \zeta^{-1} 
\rho(v_{1}) \cdots \rho(\zeta Q\cdot v_{l}) \cdots \rho(v_{k}) + 
\sum_{\alpha} c^{\alpha} 
\rho(v_{1}) \cdots \rho(\partial_{\alpha}v) \cdots \rho(v_{k}) 
\right] A(x)
\label{dQV}
\eea
\end{prop}

The proof is done elementary using the commutation relations (\ref{partial-a})
and (\ref{dQ-a}). One can write the preceding relation more compactly
introducing some notations. We denote:
\be
Q\cdot \rho(v_{1},\cdots,v_{k}) \equiv \zeta^{-1} \sum_{l=1}^{k}
\rho(v_{1},\cdots,\zeta Q\cdot v_{l},\dots,v_{k}),
\ee
and
\be
\delta\cdot \rho(v_{1},\cdots,v_{k}) \equiv \sum_{l=1}^{k} \sum_{\alpha}
c^{\alpha} \rho(v_{1},\cdots,\partial_{\alpha}v_{l},\dots,v_{k}).
\ee

We also define:
\be
D \equiv \sum_{\alpha} c^{\alpha} \partial_{\alpha}.
\ee

Then the relation from the preceding proposition can be rewritten as follows:
\be
d_{Q} \rho(V)A(x) = i (D + Q + \delta ) \rho(V)A(x)
\ee
where
$V = \{v_{1},\dots,v_{k}\}.$

For further use, we give the following commutation relations:
\be
Q\cdot \rho(v,V) - \rho(v) Q\cdot \rho(V) 
= \zeta^{-1} \rho(\zeta Q\cdot v) \rho(V)
\ee
and
\be
\delta\cdot \rho(v,V) - \rho(v) \delta\cdot \rho(V) 
= \sum_{\alpha} c^{\alpha} \rho(\partial_{\alpha}v) \rho(V).
\ee

We define a perturbation theory of the general gauge theory
$A(x)$.
We say that the  chronological products verifying, beside Bogoliubov axioms
(and other normalization conditions imposed in the analysis from the preceding
Section) verify {\it gauge invariance of rank k} if the following identity is
true for any
$|V| = k$:
\bea
\sum_{V_{1},\dots,V_{n} \in Part(V)}
\{ d_{Q} T(\rho(V_{1}) A(x_{1}),\dots,\rho(V_{k}) A(x_{k})) 
\nonumber \\
- i \sum_{l=1}^{n}
[ D_{l} \cdot T(\rho(V_{1}) A(x_{1}),\dots,\rho(V_{k}) A(x_{k}))
\nonumber \\
+ T(\rho(V_{1}) A(x_{1}),\dots,(Q + \delta)\cdot \rho(V_{l}) A(x_{l}),
\dots,\rho(V_{k}) A(x_{k}))] \} = 0;
\eea
here
\be
D_{l} \equiv \sum_{\alpha} ({\bf 1} \otimes \cdots \otimes c^{\alpha} \otimes
\cdots {\bf 1} ) \partial_{\alpha}^{l}.
\ee

It is not very hard to see that for a Yang-Mills model, the preceding relation
for 
$k = 0$
goes into the usual gauge invariance condition for the chronological product of
the Wick monomials
$A^{i}(x)$
so we call this case simply {\it gauge invariance}. The cases
$k > 0$
give the behaviour with respect to the BRST operator of the chronological
products of derivatives of the Wick monomials
$A^{i}(x)$.
There is a connection between gauge invariance of rank $k$ and rank invariance
of rank
$k + 1$
described in the following theorem which is the analogue of the result from
Appendix B of \cite{DF1}.
\begin{thm}
Suppose that
$A(x)$
verifies the gauge invariance condition of rank 
$k + 1$.
Then the anomalies of the gauge invariance condition of rank $k$ can only
appear into the vacuum sector.
\end{thm}
{\bf Proof:}
As in \cite{DF1} we consider the gauge invariance condition of rank $k$ and
commute both sides with an arbitrary
$\varphi_{w}(y)$.
Using the induction hypothesis, Wick theorem and some commutation relations
derived before we get (after some tedious but straightforward computations) the
same expression.  This means that the anomaly if proportional to {\bf 1} so it
can show up only in the vacuum sector.  
$\qed$

We say that a BRST transformation is {\it normal} if it can be expressed as the
(graded) commutator with some operator $Q$ verifying
$Q \Omega = 0, \quad Q^{*} \Omega = 0.$
The operator $Q$ is called {\it supercharge}. In usual gauge models the BRST
transformation is always normal. We have now:
\begin{cor}
If 
$d_{Q}$
is a normal BRST transformation, then the gauge invariance condition is
equivalent to the following set of identities:
\bea
\sum_{V_{1},\dots,V_{n} \in Part(V)} \{ \sum_{l=1}^{n} D_{l} 
<\Omega, T(\rho(V_{1}) A(x_{1}),\dots,\rho(V_{k}) A(x_{k}))\Omega>
\nonumber \\
+ <\Omega,T(\rho(V_{1}) A(x_{1}),\dots,(Q + \delta) \cdot \rho(V_{l}) A(x_{l}),
\dots,\rho(V_{k}) A(x_{k}))] \Omega> \} = 0
\label{Ward}
\eea
for any set of derivatives $V$.
\end{cor}

{\bf Proof:}
First we note the fact that gauge invariance of rank $k$ is always true for $k$
large enough. Indeed, the anomaly of the gauge invariance of rank $0$ is a
quasi-local operator where there is a limitation on the degree of the
polynomial in the partial derivatives - see the relation (\ref{ano}); details
of the argument can be found, for instance, in \cite{qed}. If one considers
instead of the Wick polynomials
$A^{i}(x)$
their derivatives
$\rho(V_{i})A^{i}(x)$
one can easily see that every derivative lowers the restriction on the degree
of the anomaly with at least one unit. This proves the preceding assertion. Now
the Ward identities are the vacuum averages of the gauge invariance relations
of arbitrary rank.  We apply the preceding theorem iteratively and we obtain
the conclusion.
$\qed$

It is not so simple to eliminate the anomalies from the vacuum sector.  In the
case of quantum electrodynamics \cite{DF1} this can be done using charge
conjugation invariance. In the case of a Yang-Mills model, we have some
restrictions coming from ghost number counting and PCT invariance, but they do
not eliminate all anomalies. However, we have another trick which eliminates
the anomalies in higher orders if they are absent in lower orders of
perturbation theory. We say that a gauge model
$A(x)$
is {\it of degree} $r$ if
$
c^{\alpha} = 0, \quad \forall \alpha \not= r.
$
The case considered in the Introduction corresponds to
$r = 1$ 
so we refer from now on to the notations introduced in the
formul\ae~(\ref{descent}) and (\ref{gauge-model}).
 
\begin{thm}
Let
$d_{Q}$
be a normal BRST transformation and
$A(x)$
a gauge model of degree $1$ such that
$\omega(A(x)) = d \equiv {\rm dim}(M)$
and only first order derivatives of the basic fields do appear.  Suppose that
the gauge invariance of rank $k$ is valid up to the order $k$ of the
perturbation theory
$\forall k = 0, 1,\dots, d+1$.
Then the chronological products can be chosen in such a way that the the gauge
invariance of arbitrary order $k$ is valid in every order of the perturbation
theory. 
\end{thm}
{\bf Proof:}
(i) First we prove that one can choose the chronological products in such a way
that
\be
\omega\left({\partial\over \partial x^{\mu}_{1}}
T(T^{\mu\dots}(x_{1}),A_{2}(x_{2}),\dots,A_{n}(x_{n}))\right)
= \omega\left(T(T^{\mu\dots}(x_{1}),A_{2}(x_{2}),\dots,A_{n}(x_{n}))\right) + 1
\label{increase}
\ee

Indeed, for
$n = 1$
the assertion is true because the only way to break this equality is through
the use of the equations of motion. But only Wick monomials of the type
\be
\partial^{\mu}\varphi, \quad \overline{\psi} \gamma^{\mu} \psi
\ee
(where
$\psi$
is a Dirac field) can break the preceding equality and they are of canonical
dimension
$\omega \leq 3$.
If the assertion is true for
$1, 2,\dots, n-1$
then we have a formula of this type for the corresponding commutator function
(\ref{wick-d}). A similar formula is valid for for the commutator function with
derivatives if the first factor is not differentiated. It follows that if we
consider the Wick expansion of these commutator function we will have terms for
which a formula of type (\ref{increase}) is valid for the orders of singularity
of the numerical distributions. The order of singularity is not modified by a
suitable causal distribution splitting so the formula (\ref{increase}) is
pushed to the order $n$.

(ii) Now, gauge invariance of rank $d+2$ is valid because of the restrictions
imposed on the degree anomalies: the argument was also used in the preceding
Corollary. We consider now the gauge invariance of rank $d+1$. We use induction
on the order of perturbation theory. According to the hypothesis of the
theorem the gauge condition is true up to the order $d+1$ of the perturbation
theory. We suppose that it is true up to the order $n-1$. The obstructions in
order $n$ are given by the relations (\ref{Ward}). More precisely one has to
investigate if it possible to split causally the relations:
\bea
\sum_{V_{1},\dots,V_{n} \in Part(V)} \{ \sum_{l=1}^{n} D_{l} 
<\Omega, D(\rho(V_{1}) A(x_{1}),\dots,\rho(V_{n}) A(x_{n}))\Omega>
\nonumber \\
+ <\Omega,D(\rho(V_{1}) A(x_{1}),\dots,(Q + \delta) \cdot \rho(V_{l}) A(x_{l}),
\dots,\rho(V_{n}) A(x_{n}))] \Omega> \} = 0
\label{dd}
\eea
where
$
D(A^{i_{1}}(x_{1}),\dots,A^{i_{n}}(x_{n}))
$
are the commutator distrubutions with causal support introduced in Subsection
\ref{eg}.

Because the gauge model is of degree $1$ these identities have the generic
form:
\bea
\sum_{p \in Part(V)} \sum_{l=1}^{n} 
{\partial\over \partial x^{\mu}_{l}}
d^{\mu\nu\dots}_{l;p}(x_{1},\dots,x_{n}) =
\sum_{p \in Part(V)} 
d^{\nu\dots}_{p}(x_{1},\dots,x_{n})
\label{Ward-d}
\eea
where the distributions
$
d^{\mu\nu\dots}_{l;p}(x_{1},\dots,x_{n})
$
and
$
d^{\nu\dots}_{p}(x_{1},\dots,x_{n})
$
have causal support. For certain partitions  $p$ these identities might be
purely algebraic i.e. the terms with derivatives are missing and the
equations are of the type
\bea
\sum_{p \in Part(V)} d^{\nu\dots}_{p}(x_{1},\dots,x_{n}) = 0.
\label{Ward-alg}
\eea

We call them {\it algebraic} Ward identities; the other Ward identities are
called {\it non-algebraic}. The corresponding partitions $a$ are also called 
{\it algebraic} and {\it non-algebraic} as well. In this case, we consider only
a independent set of distributions, eliminate the others algebraically and
substitute the result into the non-algebraical Ward identities (\ref{Ward-d}).
The result will be a system of equations of the same type where in all
equations the derivative terms do appear explicitly.

The resulting system equation has the following generic form: we define
\be
d^{\nu\dots}_{l;p} \equiv
{\partial \over \partial x^{\mu}_{l}} d^{\mu\nu\dots}_{l;p}
\label{ward-dd}
\ee
and we must have
\be
d^{\nu\dots}_{p} = \sum_{l=1}^{n} d^{\nu\dots}_{l;p};
\ee
here the index $p$ runs over the the set of non-algebraic partitions
${\cal P}' \subset Part(V)$.

The anomalies can appear because for some of these indices 
$p \in {\cal P}'$
we can have 
\be
\omega({\partial \over \partial x^{\mu}_{l}} d^{\mu\nu\dots}_{l;p})
< \omega(d^{\nu\dots}_{l;p}) + 1;
\ee
we denote by 
$A_{l}$ 
the set of these (non-algebraic) anomalous partitions and by
$A'_{l}$
the complement in
${\cal P}'$;
Let us observe that
$A'_{l} \not= \emptyset$.
We consider some causal splitting of the distributions from the relations
(\ref{ward-dd})
\be
d^{\nu\dots}_{l;p} = a^{\nu\dots}_{l;p} - r^{\nu\dots}_{l;p}, \quad
d^{\mu\nu\dots}_{l;p} = a^{\mu\nu\dots}_{l;p} - r^{\mu\nu\dots}_{l;p}
\ee
such that Lorentz covariance and the order of singularity is preserved. Some
anomalies might appear because of the existence of partitions of type
$A_{l}$;
let the sum of all anomalies be denote by
$P$.
There is a limitation on the degree of the anomaly, namely it cannot exceed
$supp(\omega(d_{l;p}))$.
The maximum is reached for some non-anomalous partition
$a \in A'_{l}$;
then we can add the anomaly to the expression
$a^{\nu\dots}_{l;p}$
(resp.
$r^{\nu\dots}_{l;p}$)
i.e. we can make the finite renormalization of the advanced part
$a^{\nu\dots}_{l;p} \rightarrow a^{\nu\dots}_{l;p} + P$
and similarly for the retarded part.  In this way we do not spoil Lorentz
covariance and we preserve the order of singularity. It can be seen that this
renormalization procedure can be done in a non-contradictory way in all Ward
identities: this follows from the fact that distinct Ward identities correspond
to distinct choices of the Wick polynomials
$A^{i_{p}}, \quad p =1,\dots,n$
and/or distinct sets of derivatives $V$ and we have eliminated all algebraic
constraints on the chronological products.  
$\qed$

\newpage
\section{The Gauge Invariance of the Yang-Mills Model}

So, the gauge invariance of a gauge model of degree $1$ can be reduced to the
investigation of a finite number of Ward identities. By tedious computation one
can prove that this is true for the generalization of the standard model
considered in \cite{standard} and \cite{fermi}. Let us give some details.

\subsection{Yang-Mills Fields\label{ym}}
         
In \cite{YM} - \cite{fermi} we have considered the following scheme for the
standard model (SM): we construct the auxiliary Hilbert space 
${\cal H}_{YM}^{gh,r}$
from the vacuum
$\Omega$
by applying the free fields
$
A_{a\mu},~ u_{a},~\tilde{u}_{a},~\Phi_{a}, \quad a = 1,\dots,r
$
and
$\psi_{A}, \quad A = 1,\dots,N$.
The fields
$\psi_{A}$
are, in general Dirac fields describing the matter and have the masses
$M_{A}, \quad A = 1,\dots,N$.
We give the spin structure and the statistics for the other fields: first we
postulate that
$
A_{a\mu}
$
(resp.
$
u_{a},~\tilde{u}_{a},~\Phi_{a}, \quad a = 1,\dots,r)
$
has vector (resp. scalar) transformation properties with respect to the
Poincar\'e group. In other words, every vector field has three scalar
partners.. Also
$
A_{a\mu}, ~\Phi_{a}
$
are Boson and
$
u_{a},~\tilde{u}_{a} \quad a = 1,\dots,r
$
are Fermion fields. 

Moreover,: if for some index $a$ the vector field
$A_{a}^{\mu}$ 
has non-zero mass
$m_{a}$
then we suppose that all the other scalar partners fields
$
u_{a},~\tilde{u}_{a},~\Phi_{a}
$
have the same mass $m_{a}$.

If for some index $a$ the vector field
$A_{a}^{\mu}$ 
has zero mass then the scalar partners fields
$
u_{a},~\tilde{u}_{a}
$
also have the zero mass but the corresponding scalar field 
$\Phi_{a}$
can have a arbitrary mass 
$m^{*}_{a}$
or might be absent.

Finally, we admit that for some indices $a$ all the fields
$
A_{a}^{\mu}, u_{a},~\tilde{u}_{a}
$
might be absent and the corresponding scalar field 
$\Phi_{a}$
can have a arbitrary mass
$m^{*}_{a}$.

The canonical (anti)commutation relations are:
\bea
\left[A_{a\mu}(x),A_{b\nu}(y)\right] = - 
\delta_{ab} g_{\mu\nu} D_{m_{a}}(x-y) \times {\bf 1},
\nonumber \\
\{u_{a}(x),\tilde{u}_{b}(y)\} = \delta_{ab} D_{m_{a}}(x-y) \times {\bf 1}, 
\quad
[ \Phi_{a}(x),\Phi_{b}(y) ] = \delta_{ab} D_{m^{*}_{a}}(x-y) \times {\bf 1};
\nonumber \\
\{\psi_{A}(x),\overline{\psi_{B}}(y)\} = \delta_{AB} S_{M_{A}}(x-y)
\label{comm-r}
\eea
all other (anti)commutators are null.
                        
In the Hilbert space 
${\cal H}_{YM}^{gh,r}$
we suppose given a sesquilinear form 
$<\cdot, \cdot>$
such that:
\be
A_{a\mu}(x)^{\dagger} = A_{a\mu}(x), \quad
u_{a}(x)^{\dagger} = u_{a}(x), \quad
\tilde{u}_{a}(x)^{\dagger} = - \tilde{u}_{a}(x), \quad
\Phi_{a}(x)^{\dagger} = \Phi_{a}(x).
\label{conjugate-YM}
\ee
                        
The ghost degree is 
$\pm 1$ 
for the fields
$~u_{a}$
(resp.
$
~\tilde{u}_{a}),
\quad a = 1,\dots,r
$
and $0$ for the other fields.

One can define the BRST {\it supercharge} $Q$ by:
\bea
\{Q, u_{a} \}= 0 \quad 
\{Q, \tilde{u}_{a} \}= - i (\partial_{\mu} A_{a}^{\mu} + m_{a} \Phi_{a})
\nonumber \\
~[ Q, A_{a}^{\mu} ] = i \partial^{\mu} u_{a} \quad
[ Q, \Phi_{a} ] = i m_{a} u_{a}, \quad \forall a = 1,\dots,r
\label{BRST-YM}
\eea
and 
\be
Q \Omega = 0.
\ee

Then one can justify that the {\bf physical} Hilbert space of the Yang-Mills
system is a factor space
\be
{\cal H}^{r}_{YM} \equiv {\cal H} \equiv Ker(Q)/Ran(Q).
\ee

The sesquilinear form
$<\cdot, \cdot>$
induces a {\it bona fide} scalar product on the Hilbert factor space.

Let us consider the set of Wick monomials ${\cal W}$ constructed from the free
fields
$A_{a}^{\mu},~u_{a},~\tilde{u}_{a}$
and
$\Phi_{a}$
for all indices
$a = 1,\dots,r$;
we define the BRST operator
$d_{Q}: {\cal W} \rightarrow {\cal W}$
as the (graded) commutator with the supercharge operator $Q$. Then one can
prove easily that:
\be
d_{Q}^{2} = 0.
\label{coomology}
\ee

Let us consider the first order Lagrangian:
\bea
T(x) \equiv
f_{abc} \left[ {1\over 2} :A_{a\mu}(x)A_{b\nu}(x) F_{a}^{\mu\nu}(x):
- :A_{a}^{\mu}(x) u_{b}(x) \partial_{\mu} \tilde{u}_{c}(x):\right]
\nonumber \\
+ f'_{abc} \left[ :\Phi_{a}(x) \partial_{\mu} \Phi_{b}(x) A_{c}^{\mu}(x): 
- m_{b} :\Phi_{a}(x) A_{b\mu}(x) A_{c}^{\mu}(x): 
- m_{b} :\Phi_{a}(x) \tilde{u}_{b}(x) u_{c}(x):\right]
\nonumber \\
+ f^{"}_{abc} :\Phi_{a}(x) \Phi_{b}(x) \Phi_{c}(x): 
+  j^{\mu}_{a}(x) A_{a\mu}(x) + j_{a}(x) \Phi_{a}(x)
\label{inter}
\eea
where:
\be
F_{a}^{\mu\nu}(x) \equiv 
\partial^{\mu} A^{\nu}_{a}(x) - \partial^{\nu} A^{\mu}_{a}(x) 
\ee
is the Yang-Mills field tensor and the so-called {\it currents} are:
\be
j_{a}^{\mu}(x) = 
:\overline{\psi_{A}}(x) (t_{a})_{AB} \gamma^{\mu} \psi_{B}(x): +
:\overline{\psi_{A}}(x) (t'_{a})_{AB} \gamma^{\mu} \gamma_{5} \psi_{B}(x):
\label{vector-current}
\ee
and
\be
j_{a}(x) = 
:\overline{\psi_{A}}(x) (s_{a})_{AB} \psi_{B}(x): +
:\overline{\psi_{A}}(x) (s'_{a})_{AB} \gamma_{5} \psi_{B}(x):
\label{scalar-current}
\ee
where a number of restrictions must be imposed on the various constants (see
\cite{YM}-\cite{fermi}.

Moreover, if we define
\bea
T^{\mu}(x)  = f_{abc} \left[ :u_{a}(x) A_{b\nu}(x) F^{\nu\mu}_{c}(x): -
{1\over 2} :u_{a}(x) u_{b}(x) \partial^{\mu}(x) \tilde{u}_{c}(x): \right]
\nonumber \\
+ f'_{abc} \left[ m_{a} :A_{a}^{\mu}(x) \Phi_{b}(x) u_{c}(x):
+ :\Phi_{a}(x) \partial^{\mu}\Phi_{b}(x) u_{c}(x): \right].
+  u_{a}(x) j^{\mu}_{a}(x)
\label{inter-mu}
\eea
and
\be
T^{\mu\nu}(x)  = {1\over 2} f_{abc} :u_{a}(x) u_{b}(x) F^{\nu\mu}_{c}(x):
\ee
then we have the relation (\ref{descent}) from the Introduction for
$p =2$.

All these Wick polynomials are
$SL(2,\C)$-covariant, are causally commuting and are Hermitean. Moreover we
have the following ghost content:
\be
gh(T(x)) = 0, \quad gh(T^{\mu}(x)) = 1, \quad gh(T^{\mu\nu}(x)) = 2.
\ee

We will construct a perturbation theory verifying Bogoliubov axioms using this
set of free fields and imposing the usual axioms of causality, unitarity and
relativistic invariance on the chronological products
$T(A^{i_{1}}(x_{1}),\dots,A^{i_{n}}(x_{n}))$
(where the Wick polynomials
$A^{i}(x)$
must be
$T(x), \quad T^{\mu}(x)$
or
$T^{\mu\nu}(x)$)
such that we have the relation (\ref{gauge-model}) from the Introduction which
amounts to some factorizes property of the chronological products to the
physical Hilbert space in the formal adiabatic limit. This generalizes the
gauge invariance condition from \cite{AS}, \cite{DS}:
\be
d_{Q} T(T(x_{1}),\dots,T(x_{n})) = i \sum_{l=1}^{n} 
{\partial \over \partial x^{\mu}_{l}} 
T(T(x_{1}),\dots,T^{\mu}_{l}(x_{l}),\dots,T(x_{n})).
\label{gauge-inf}
\ee

We work from now on with the usual chronological products. The various signs
from som of the relations below are obtained by conveniently eliminating the
Grassmann variables.

Let us consider now some elements
$v_{1},\dots,v_{k} \in {\cal M}^{0}$
of fixed ghost number and let us define:
\be
g_{l} = \sum_{i=1}^{l-1} gh(v_{i}), \quad
g'_{l} = \sum_{i=l}^{k} gh(v_{i}).
\ee
Then after some computation one obtains from (\ref{dQV}):
\bea
d_{Q} [ \rho(v_{1})\dots \rho(v_{k}) T(x) ] =
i \partial_{\mu} [\rho(v_{1},\dots,\rho(v_{k}) T^{\mu}(x)]
\nonumber \\
+ i \sum_{l=1}^{k} [(-1)^{g_{l}} 
\rho(v_{1}) \dots,\rho(q\cdot v_{l}) \rho(v_{k}) T(x)
+ \rho(v_{1}) \dots,\rho(\partial_{\mu}\cdot v_{l}) \rho(v_{k}) T^{\mu}(x) ]
\label{Ward1}
\eea
\bea
d_{Q} [ \rho(v_{1})\dots \rho(v_{k}) T^{\mu}(x) ] =
i \partial_{\nu} [\rho(v_{1},\dots,\rho(v_{k}) T^{\nu\mu}(x)]
\nonumber \\
+ i \sum_{l=1}^{k} [(-1)^{g'_{l}}
\rho(v_{1}) \dots,\rho(q\cdot v_{l}) \rho(v_{k}) T^{\mu}(x)
+ \rho(v_{1}) \dots,\rho(\partial_{\nu}\cdot v_{l}) \rho(v_{k}) T^{\nu\mu}(x) ]
\label{Ward2}
\eea
\bea
d_{Q} [ \rho(v_{1})\dots \rho(v_{k}) T^{\mu\nu}(x) ] =
i \sum_{l=1}^{k} (-1)^{g_{l}} \rho(v_{1}) \dots,\rho(q\cdot v_{l}) \rho(v_{k})
T^{\mu\mu}(x).
\label{Ward3}
\eea

Here the expressions
$\partial_{\mu}\cdot v$
are defined according to (\ref{der}) and
\bea
q\cdot {\partial \over \partial A_{a\mu}} = - {1 \over 4} m_{a}^{2}
{\partial \over \partial \tilde{u}_{a;\mu}}, \quad
q\cdot {\partial \over \partial u_{a}} = {1 \over 4} m_{a}^{2} g_{\mu\nu}
{\partial \over \partial A_{a\mu;\nu}}, \quad
q\cdot {\partial \over \partial \tilde{u}_{a}} = 0, \quad
q\cdot {\partial \over \partial \Phi_{a}} = m_{a}
{\partial \over \partial \tilde{u}_{a}},
\nonumber \\
q\cdot {\partial \over \partial A_{a\mu;\nu}} = g^{\mu\rho}
{\partial \over \partial \tilde{u}_{a}}, \quad
q\cdot {\partial \over \partial u_{a;\mu}} = 
- {\partial \over \partial A_{a\mu}}, \quad
q\cdot {\partial \over \partial \tilde{u}_{a;\mu}} = 0, \quad
q\cdot {\partial \over \partial \Phi_{a;\mu}} = m_{a}
{\partial \over \partial \tilde{u}_{a;\mu}}; \qquad
\eea
where the derivatives with respect to the fields are defined according to the
general formul\ae~ (\ref{der-A}). 

Using these relations one can easily write now explicitly all Ward identities.
We will not list them here.
\newpage

\subsection{Lower Order Ward Identities}

We consider the identities (\ref{Ward}) for
$2 \leq n \leq 5$.
It follows from the preceding Section that we have have something non-trivial 
only if
\be
|V| \leq 5, \quad 
gh(V) = \sum_{l=1}^{n} gh(A^{i_{l}}) - 1.
\ee

Also, in the sum over the partitions of $V$ it is sufficient to consider only
those terms for which all subsets
$V_{1},\dots,V_{n}$
are non-void. The other partitions are not dangerous: they can be treated as in
the last theorem of the preceding Section. Finally, at least one term from 
(\ref{Ward}) which is differentiated should be of the form
\be
{\partial \over \partial x^{\mu}_{l}} T(\dots,\partial^{\mu}\psi(x_{l}),\dots),
\quad
{\partial \over \partial x^{\mu}_{l}} 
T(\dots,\overline{\psi}(x_{l})\gamma^{\mu}\psi(x_{l}),\dots)
\label{factor}
\ee
for
$\psi$
a Dirac field. Only in this cases we will have a anomalous partition and some
anomaly might appear.

The list of these Ward identities is too long to give in detail. We will only
mention the choices for the set $V$ and insist on those identities which are
producing anomalies. Afterwards we will specify the finite renormalizations
which do eliminate the anomalies. In all these computations we heavily relay on
the various relations verified by the constants appearing in the first order
Lagrangian 
$T(x)$;
all these constraints can be found in \cite{standard} and \cite{fermi}.

(i) $n = 2$

In the second order perturbation theory we have three possibilities 

(i1) $A^{i_{1}}(x) = A^{i_{2}}(x) = T(x)$

In this case we must have
$gh(V) = 1$
so we have the following cases:
\bea
V = \left\{{\partial \over \partial u_{a}}, {\partial \over \partial u_{b}}, 
{\partial \over \partial \tilde{u}_{c}}, {\partial \over \partial A_{d\rho}}
\right\}, \quad
V = \left\{{\partial \over \partial u_{a}}, {\partial \over \partial u_{b}}, 
{\partial \over \partial \tilde{u}_{c}}, {\partial \over \partial \Phi_{d}}
\right\},
\nonumber \\
V = \left\{{\partial \over \partial u_{a}}, 
{\partial \over \partial A_{b\nu}}, 
{\partial \over \partial A_{c\rho}}, {\partial \over \partial A_{d\sigma}} 
\right\}, \quad
V = \left\{{\partial \over \partial u_{a}}, 
{\partial \over \partial A_{b\nu}}, 
{\partial \over \partial A_{c\rho}}, {\partial \over \partial \Phi_{d}} 
\right\},
\nonumber \\
V = \left\{{\partial \over \partial u_{a}}, 
{\partial \over \partial A_{b\nu}}, 
{\partial \over \partial \Phi_{c}}, {\partial \over \partial \Phi_{d}} 
\right\}, \quad
V = \left\{{\partial \over \partial u_{a}}, 
{\partial \over \partial \Phi_{b}}, 
{\partial \over \partial \Phi_{c}}, {\partial \over \partial \Phi_{d}} 
\right\}
\nonumber \\
V = \left\{{\partial \over \partial u_{a}}, 
{\partial \over \partial \Phi_{b}}, 
{\partial \over \partial \Phi_{c}}, {\partial \over \partial \Phi_{d}},
{\partial \over \partial \Phi_{e}} \right\}, 
\quad
V = \left\{{\partial \over \partial u_{a}}, 
{\partial \over \partial A_{b\rho}} 
\right\},
\quad
V = \left\{{\partial \over \partial u_{a}}, {\partial \over \partial \Phi_{b}} 
\right\}.
\eea
and the cases obtained from the first six ones by appending a derivatives to one
of the fields. In all, there are 24 such relations. We give below only the
anomalous Ward identities:

$$ 1) \quad
V = \left\{{\partial \over \partial u_{a}}, 
{\partial \over \partial A_{b\nu}}, 
{\partial \over \partial A_{c\rho}}, 
{\partial \over \partial A_{d\sigma;\lambda}} 
\right\}
$$

Let us give the Ward identity in detail in the case:
\bea
{\partial \over \partial x^{\mu}_{1}}
<\Omega, T\left({\partial^{2} \over \partial u_{a} \partial A_{b\nu}} 
T^{\mu}(x_{1}),
{\partial^{2} \over \partial A_{c\rho} \partial A_{d\sigma;\lambda}} T(x_{2})
\right)\Omega>
\nonumber \\
+ <\Omega, T\left({\partial^{2} \over \partial u_{a} \partial A_{d\sigma}} 
T^{\lambda}(x_{1}),
{\partial^{2} \over \partial A_{b\nu} \partial A_{c\rho}} T(x_{2})
\right)\Omega>
\nonumber \\
+ (b\nu \leftrightarrow c\rho) + (x_{1} \leftrightarrow x_{2}) + \cdots = 0
\eea
where by $\dots$ we mean terms which do not produce anomalies. 

We will illustrate the procedure of getting the anomaly on this case. The
first chronological product comes from the causal commutator
\be
\left[ {\partial^{2} \over \partial u_{a} \partial A_{b\nu}} 
T^{\mu}(x_{1}),
{\partial^{2} \over \partial A_{c\rho} \partial A_{d\sigma;\lambda}} T(x_{2})
\right] = f_{abe} f_{cde} 
(g^{\rho\sigma} g^{\nu\lambda} - g^{\rho\lambda} g^{\nu\sigma})
\partial^{\mu} D_{m_{e}}(x_{1}-x_{2})
\ee
and it produces the anomaly
\be
f_{abe} f_{cde} (g^{\rho\sigma} g^{\nu\lambda} - g^{\rho\lambda} g^{\nu\sigma})
\delta(x_{1}-x_{2}).
\ee

The total anomaly produced by the preceding Ward identity is
\be
A_{1;abcd}^{\nu\rho\sigma\lambda} = 2i f_{ade} f_{cbe} 
(g^{\rho\sigma} g^{\nu\lambda} - g^{\rho\lambda} g^{\nu\sigma})
\delta(x_{1}-x_{2}).
\ee

$$ 2) \quad
V = \left\{{\partial \over \partial u_{a}}, {\partial \over \partial A_{b\nu}}, 
{\partial \over \partial A_{c\rho}}, 
{\partial \over \partial A_{d\sigma}} \right\}
$$
with the anomaly
\be
A_{2;abcd}^{\nu\rho\sigma} = i f_{abe} f_{cde} 
(g^{\nu\sigma} \partial^{\rho} - g^{\nu\rho} \partial^{\sigma})
\delta(x_{1}-x_{2}) 
+ (b\nu \leftrightarrow c\rho) + (b\nu \leftrightarrow d\sigma)
\ee

$$ 3) \quad
V = \left\{{\partial \over \partial u_{a;\nu}}, 
{\partial \over \partial A_{b\rho}}, 
{\partial \over \partial A_{c\sigma}}, 
{\partial \over \partial A_{d\lambda}} \right\}
$$

In this case we get a algebraic Ward identity:
\bea
<\Omega, T\left({\partial^{2} \over \partial A_{a\nu} \partial A_{b\rho}} 
T(x_{1}),
{\partial^{2} \over \partial A_{c\sigma} \partial A_{d\lambda\rho}} 
T(x_{2})\right)\Omega>
\nonumber \\
- <\Omega, T\left({\partial^{2} \over \partial u_{a} \partial A_{b\rho}} 
T^{\nu}(x_{1}),
{\partial^{2} \over \partial A_{c\sigma} \partial A_{d\lambda\rho}} 
T(x_{2})\right)\Omega>
\nonumber \\
+ (b\rho \leftrightarrow c\sigma) + (b\rho \leftrightarrow d\lambda) 
+ (x_{1} \leftrightarrow x_{2}) + \cdots = 0
\eea

$$ 4) \quad
V = \left\{{\partial \over \partial u_{a}}, 
{\partial \over \partial A_{b\rho}}, 
{\partial \over \partial \Phi_{c}}, {\partial \over \partial \Phi_{d;\sigma}} 
\right\}
$$
with the anomaly
\be
A_{3;abcd}^{\rho\sigma} 
= - 2i g^{\rho\sigma} f'_{dea} f'_{ceb} \delta(x_{1}-x_{2}) 
\ee

$$ 5) \quad
V = \left\{{\partial \over \partial u_{a}}, 
{\partial \over \partial A_{b\rho}}, 
{\partial \over \partial \Phi_{c}}, {\partial \over \partial \Phi_{d}} 
\right\}
$$
with the anomaly
\be
A_{4;abcd}^{\rho} = - 2i ( f'_{dea} f'_{ceb} + f'_{deb} f'_{cea})
\partial^{\rho}\delta(x_{1}-x_{2}) 
\ee

$$ 6) \quad
V = \left\{{\partial \over \partial u_{a;\rho}}, 
{\partial \over \partial A_{b\sigma}}, 
{\partial \over \partial \Phi_{c}}, {\partial \over \partial \Phi_{d}} \right\}
$$
In this case we get a algebraic Ward identity.

$$ 7) \quad
V = \left\{{\partial \over \partial u_{a}}, 
{\partial \over \partial \Phi_{b}}, 
{\partial \over \partial \Phi_{c}}, {\partial \over \partial \Phi_{d}} \right\}
$$
with the anomaly
\be
A_{5;abcd} = - 12i {\cal S}_{bcd}( f'_{bea} f"_{cde} ) \delta(x_{1}-x_{2}) 
\ee

(i2) $A^{i_{1}}(x) = T(x), \quad A^{i_{2}}(x) = T^{\nu}(x)$

In this case we must have
$gh(V) = 2$
so we have the following possibilities:
\bea
V = \left\{{\partial \over \partial u_{a}}, {\partial \over \partial u_{b}}, 
{\partial \over \partial u_{c}}, {\partial \over \partial \tilde{u}_{d}}
\right\}, \quad
V = \left\{{\partial \over \partial u_{a}}, {\partial \over \partial u_{b}}, 
{\partial \over \partial A_{c\rho}}, {\partial \over \partial A_{d\sigma}}
\right\},
\nonumber \\
V = \left\{{\partial \over \partial u_{a}}, {\partial \over \partial u_{b}}, 
{\partial \over \partial A_{c\rho}}, {\partial \over \partial \Phi_{d}} 
\right\}, \quad
V = \left\{{\partial \over \partial u_{a}}, {\partial \over \partial u_{b}}, 
{\partial \over \partial \Phi_{c}}, {\partial \over \partial \Phi_{d}} 
\right\}, \quad
V = \left\{{\partial \over \partial u_{a}}, 
{\partial \over \partial u_{b}}, \right\}, 
\eea
and the cases obtained from the first four ones by appending a derivatives to
one of the fields. In all, there are 14 such relations. We give below only the
anomalous Ward identities:

$$ 8) \quad
V = \left\{{\partial \over \partial u_{a}}, {\partial \over \partial u_{b}}, 
{\partial \over \partial A_{c\rho}}, 
{\partial \over \partial A_{d\sigma;\lambda}} \right\}
$$
with the anomaly
\be
A_{6;abcd}^{\nu\rho\sigma\lambda} = - i f_{abe} f_{cde} 
(g^{\rho\sigma} g^{\nu\lambda} - g^{\rho\lambda} g^{\nu\sigma})
\delta(x_{1}-x_{2})
\ee

$$ 9) \quad
V = \left\{{\partial \over \partial u_{a}}, {\partial \over \partial u_{b}}, 
{\partial \over \partial A_{c\rho}}, 
{\partial \over \partial A_{d\sigma}} \right\}
$$
with the anomaly
\be
A_{7;abcd}^{\nu\rho\sigma} = 2i f_{abe} f_{cde} 
(g^{\nu\sigma} \partial^{\rho} - g^{\nu\rho} \partial^{\sigma})
\delta(x_{1}-x_{2}) 
+ (b\nu \leftrightarrow c\rho) + (b\nu \leftrightarrow d\sigma)
\ee

$$ 10) \quad
V = \left\{{\partial \over \partial u_{a;\nu}}, 
{\partial \over \partial u_{b}}, 
{\partial \over \partial A_{c\sigma}}, 
{\partial \over \partial A_{d\lambda}} \right\}
$$
In this case we get a algebraic Ward identity;

$$ 11) \quad
V = \left\{{\partial \over \partial u_{a}}, {\partial \over \partial u_{b}}, 
{\partial \over \partial \Phi_{c}}, {\partial \over \partial A_{d\rho}} 
\right\}
$$
with the anomaly
\be
A_{8;abcd}^{\rho\sigma} 
= - i g^{\rho\sigma} [ m_{a} ( f'_{ead} f'_{ceb} + f'_{ced} f'_{eab})
- (a \leftrightarrow b) ] \delta(x_{1}-x_{2}) 
\ee

$$ 12) \quad
V = \left\{{\partial \over \partial u_{a}}, {\partial \over \partial u_{b}}, 
{\partial \over \partial \Phi_{c}}, {\partial \over \partial \Phi_{d;\rho}} 
\right\}
$$
with the anomaly
\be
A_{9;abcd}^{\rho} = - i g^{\rho\sigma} f_{abe} f'_{cde} \delta(x_{1}-x_{2}) 
\ee

$$ 13) \quad
V = \left\{{\partial \over \partial u_{a;\rho}}, 
{\partial \over \partial u_{b}}, 
{\partial \over \partial \Phi_{c}}, {\partial \over \partial \Phi_{d}} \right\}
$$
In this case we get a algebraic Ward identity.

(i3) $A^{i_{1}}(x) = T(x), \quad A^{i_{2}}(x) = T^{\nu\rho}(x)$

In this case we must have
$gh(V) = 3$
so we have the following possibilities:
\be
V = \left\{ {\partial \over \partial u_{a}}, {\partial \over \partial u_{b}}, 
{\partial \over \partial u_{c}}, {\partial \over \partial A_{d\rho}}
\right\}, \quad
V = \left\{ {\partial \over \partial u_{a}}, {\partial \over \partial u_{b}}, 
{\partial \over \partial u_{c}}, {\partial \over \partial \Phi_{d}}
\right\}
\ee
and the cases obtained by appending a derivatives to one of the fields. There
are 6 such relations and the corresponding Ward identities do not give
anomalies.

(i4) $A^{i_{1}}(x) = T(x), \quad A^{i_{2}}(x) = T^{\nu\rho}(x)$

In this case we also have
$gh(V) = 3$
so we have the same possibilities as in case (i3):
\bea
V = \left\{{\partial \over \partial u_{a}}, {\partial \over \partial u_{b}}, 
{\partial \over \partial u_{c}}, {\partial \over \partial A_{d\rho}}
\right\}, \quad
V = \left\{{\partial \over \partial u_{a}}, {\partial \over \partial u_{b}}, 
{\partial \over \partial u_{c}}, {\partial \over \partial \Phi_{d}}
\right\}
\eea
and the cases obtained by appending a derivatives to one of the fields. There
are 6 such relations. The anomalous Ward identities correspond to:

$$ 14) \quad
V = \left\{{\partial \over \partial u_{a;\sigma}}, 
{\partial \over \partial u_{b}}, {\partial \over \partial u_{c}}, 
{\partial \over \partial A_{d\sigma}} \right\};
$$
In this case we get a algebraic Ward identity;

$$ 15) \quad
V = \left\{{\partial \over \partial u_{a}}, {\partial \over \partial u_{b}}, 
{\partial \over \partial u_{c}}, {\partial \over \partial \Phi_{d}} \right\}
$$
with the anomaly
\be
A_{10;abcd}^{\rho\sigma} 
= - i g^{\rho\sigma} {\cal A}_{abc} (f_{abc} f'_{dec}) m_{e} 
\delta(x_{1}-x_{2}) 
\ee

(i5) $A^{i_{1}}(x) = T^{\nu}(x), \quad A^{i_{2}}(x) = T^{\rho\sigma}(x)$

In this case we take
$gh(V) = 4$
so we have only:
\be
V = \left\{{\partial \over \partial u_{a}}, {\partial \over \partial u_{b}}, 
{\partial \over \partial u_{c}}, {\partial \over \partial u_{d}}\right\}
\ee
and the case obtained by appending a derivatives to one of the fields. There
are 2 such relations and the corresponding Ward identities do not produce
anomalies. 

All anomalies can be removed if we perform the following finite renormalization
of the chronological products:
\bea
T\left({\partial^{2} \over \partial u_{a} \partial A_{d\sigma}} 
T^{\lambda}(x_{1}),
{\partial^{2} \over \partial A_{b\nu} \partial A_{c\rho}} T(x_{2})
\right) \rightarrow \cdots
+ f_{ade} f_{bce} 
(g^{\rho\sigma} g^{\nu\lambda} - g^{\rho\lambda} g^{\nu\sigma})
\delta(x_{1}-x_{2})
\nonumber \\
T\left({\partial^{2} \over \partial A_{a\nu} \partial A_{b\rho}} 
T(x_{1}),
{\partial^{2} \over \partial A_{c\sigma} \partial A_{d\lambda}} T(x_{2})
\right) \rightarrow \cdots
- f_{abe} f_{cde} 
(g^{\rho\sigma} g^{\nu\lambda} - g^{\rho\lambda} g^{\nu\sigma})
\delta(x_{1}-x_{2})
\nonumber \\
T\left({\partial^{2} \over \partial u_{a} \partial \Phi_{d}} 
T^{\sigma}(x_{1}),
{\partial^{2} \over \partial A_{b\rho} \partial \Phi_{c}} T(x_{2})
\right) \rightarrow \cdots
+ f'_{ceb} f'_{dea} 
(g^{\rho\sigma} g^{\nu\lambda} - g^{\rho\lambda} g^{\nu\sigma})
\delta(x_{1}-x_{2})
\nonumber \\
T\left({\partial^{2} \over \partial A_{a\rho} \partial \Phi_{c}} 
T(x_{1}),
{\partial^{2} \over \partial A_{b\sigma} \partial \Phi_{d}} T(x_{2})
\right) \rightarrow \cdots
- f'_{ceb} f'_{dea} g^{\rho\sigma} \delta(x_{1}-x_{2})
\nonumber \\
T\left({\partial^{2} \over \partial \Phi_{a} \partial \Phi_{b}} 
T(x_{1}),
{\partial^{2} \over \partial \Phi_{c} \partial \Phi_{d}} T(x_{2})
\right) \rightarrow \cdots
+ {1 \over4} {\cal S}_{bcd} ( f'_{bea} f"_{cde} ) \delta(x_{1}-x_{2})
\nonumber \\
T\left({\partial^{2} \over \partial u_{a} \partial A_{d\sigma}} 
T^{\lambda}(x_{1}),
{\partial^{2} \over \partial u_{b} \partial A_{c\rho}} T(x_{2})
\right) \rightarrow \cdots
- f_{ace} f_{bde} 
(g^{\rho\sigma} g^{\nu\lambda} - g^{\rho\lambda} g^{\nu\sigma})
\delta(x_{1}-x_{2})
\nonumber \\
T\left({\partial^{2} \over \partial A_{c\rho} \partial A_{d\sigma}} 
T(x_{1}),
{\partial^{2} \over \partial u_{a} \partial u_{b}} T^{\mu\nu}(x_{2})
\right) \rightarrow \cdots
+ f_{abe} f_{cde} 
(g^{\mu\sigma} g^{\nu\rho} - g^{\mu\rho} g^{\nu\sigma})
\delta(x_{1}-x_{2})
\nonumber \\
T\left({\partial^{2} \over \partial \Phi_{a} \partial u_{b}} 
T^{\lambda}(x_{1}),
{\partial^{2} \over \partial \Phi_{c} \partial A_{d\rho}} T^{\nu}(x_{2})
\right) \rightarrow \cdots
- f'_{ced} f'_{eab} g^{\nu\rho} \delta(x_{1}-x_{2})
\nonumber \\
T\left({\partial^{2} \over \partial u_{a} \partial \Phi_{d}} 
T^{\rho}(x_{1}),
{\partial^{2} \over \partial u_{b} \partial \Phi_{c}} T^{\nu}(x_{2})
\right) \rightarrow \cdots
+ f'_{cea} f'_{deb} g^{\rho\sigma} \delta(x_{1}-x_{2}). \qquad
\label{ren2}
\eea

All these renormalizations are made in the vacuum sector so there is no need to
take the vacuum average.  Let us note that all these finite renormalization are
consistent with the symmetry properties of the chronological products. If we
use the formula (\ref{wick-chrono2}) we can obtain the finite renormalizations
for the original chronological products:
\bea
T(T(x_{1}), T(x_{2})) \rightarrow \cdots
+ N(x_{1}) \delta(x_{1}-x_{2})
\nonumber \\
T(T^{\mu}(x_{1}), T(x_{2})) \rightarrow \cdots
+ N^{\mu}(x_{1}) \delta(x_{1}-x_{2})
\nonumber \\
T(T^{\mu}(x_{1}), T^{\nu}(x_{2})) \rightarrow \cdots
+ N^{\mu\nu}(x_{1}) \delta(x_{1}-x_{2})
\nonumber \\
T(T(x_{1}), T^{\mu\nu}(x_{2})) \rightarrow \cdots
+ N^{\mu\nu}(x_{1}) \delta(x_{1}-x_{2})
\label{ren2a}
\eea
where:
\bea
N \equiv {1\over 4} 
f_{abe} f_{cde} : A_{a\mu} A_{c}^{\mu} A_{b\rho} A_{d}^{\rho}:
+ {1\over 2} f'_{ceb} f_{dea} : A_{a\mu} A_{b}^{\mu} \Phi_{c} \Phi_{d}:
+ {1\over 2m_{a}} f'_{bea} f"_{cde} : \Phi_{a} \Phi_{b} \Phi_{c} \Phi_{d}: 
\nonumber \\
N^{\mu} \equiv
- f_{ade} f_{bce} : u_{a} A_{b}^{\mu} A_{c\rho} A_{d}^{\rho}:
- f'_{ebb} f_{eda} : u_{a} A_{b}^{\mu} \Phi_{c} \Phi_{d}:
\nonumber \\
N^{\mu\nu} \equiv
f_{abe} f_{cde} : u_{a} u_{b} A_{c\mu} A_{d\nu}: \qquad
\label{ren2b}
\eea

(ii) $n = 3$

The situation in the third order of the perturbation theory can be analysed as
in \cite{fermi}. One can see that only in two situations  anomalies can appear:

(ii2) When the chronological products involve at least one Fermionic loop. The
relevant choices for the set $V$ are
\be
V = \left\{ {\partial \over \partial u_{a}}, 
{\partial \over \partial A_{b\nu}}, 
{\partial \over \partial A_{c\rho}}, \right\}, \quad
V = \left\{ {\partial \over \partial u_{a}}, 
{\partial \over \partial A_{b\nu}}, 
{\partial \over \partial \Phi_{c}}, \right\}, \quad
V = \left\{ {\partial \over \partial u_{a}}, 
{\partial \over \partial \Phi_{b}}, 
{\partial \over \partial \Phi_{c}}, \right\}
\ee
and other relations with one of the fields differentiated. There are 10
relations of this type. The Ward identities which can produce anomalies
correspond only to the choices without derivatives. They are respectively:

- for
$A^{i_{1}}(x) =  A^{i_{2}}(x) = A^{i_{3}}(x) = T(x)$
\bea
{\partial \over \partial x^{\mu}_{1}} 
< \Omega, T(j^{\mu}_{a}(x_{1}),j^{\nu}_{b}(x_{2}),j^{\rho}_{c}(x_{3}))  
\Omega>
- m_{a} < \Omega, T(j_{a}(x_{1}),j^{\nu}_{b}(x_{2}),j^{\rho}_{c}(x_{3}))  
\Omega> 
\nonumber \\
+ (b\nu \leftrightarrow c\rho) 
+ ( x_{1} \leftrightarrow x_{2})+ ( x_{1} \leftrightarrow x_{3}) + \cdots
\eea
\bea
{\partial \over \partial x^{\mu}_{1}} 
< \Omega, T(j^{\mu}_{a}(x_{1}),j^{\nu}_{b}(x_{2}),j_{c}(x_{3}))  \Omega>
- m_{a} < \Omega, T(j_{a}(x_{1}),j^{\nu}_{b}(x_{2}),j_{c}(x_{3})) \Omega> 
\nonumber \\
+ ( x_{1} \leftrightarrow x_{2})+ ( x_{1} \leftrightarrow x_{3}) + \cdots
\eea
\bea
{\partial \over \partial x^{\mu}_{1}} 
< \Omega, T(j^{\mu}_{a}(x_{1}),j_{b}(x_{2}),j_{c}(x_{3}))  \Omega>
- m_{a} < \Omega, T(j_{a}(x_{1}),j_{b}(x_{2}),j_{c}(x_{3})) \Omega> 
\nonumber \\
+ (b \leftrightarrow c) 
+ ( x_{1} \leftrightarrow x_{2})+ ( x_{1} \leftrightarrow x_{3}) + \cdots
\eea

- for 
$A^{i_{1}}(x) =  T^{\nu}(x), A^{i_{2}}(x) = A^{i_{3}}(x) = T(x)$
\bea
{\partial \over \partial x^{\mu}_{1}} 
< \Omega, T(j^{\nu}_{a}(x_{1}),j^{\mu}_{b}(x_{2}),j^{\rho}_{c}(x_{3}))  
\Omega>
- m_{b} < \Omega, T(j^{\nu}_{a}(x_{1}),j_{b}(x_{2}),j^{\rho}_{c}(x_{3}))  
\Omega> 
\nonumber \\
- (a \leftrightarrow b) + ( x_{2} \leftrightarrow x_{3}) + \cdots
\eea

- for 
$A^{i_{1}}(x) =  T^{\nu}(x), A^{i_{2}}(x) T^{\rho}(x), A^{i_{3}}(x) = T(x)$
\bea
{\cal A}_{abc} {\partial \over \partial x^{\mu}_{1}} 
< \Omega, T(j^{\nu}_{a}(x_{1}),j^{\rho}_{b}(x_{2}),j^{\mu}_{c}(x_{3}))  
\Omega>
- m_{c} < \Omega, T(j^{\nu}_{a}(x_{1}),j^{\rho}_{b}(x_{2}),j_{c}(x_{3}))  
\Omega> 
\nonumber \\
+ ( x_{1} \leftrightarrow x_{2}) + ( x_{2} \leftrightarrow x_{3}) + \cdots
\eea

One can show as in \cite{fermi} that these Ward identities are not anomalous
if the axial vertex anomaly vanishes. Indeed, one ca show that the preceding
Ward identities can be fulfilled if the eqs. (5.1.49) - (5.1.60) from
\cite{fermi} can be causally split; this in turns happens {\it iff} the axial
anomaly vanishes.

(ii2) We also have some Ward identities where anomalies can appear because of
the finite renormalizations (\ref{ren2a}). One can easily see that these cases
correspond to the choice

$A^{i_{1}}(x) =  A^{i_{2}}(x) = A^{i_{3}}(x) = T(x)$

and the following assignments for the derivatives $V$:
\bea
V = \left\{{\partial \over \partial u_{a}}, 
{\partial \over \partial A_{b\rho}}, {\partial \over \partial A_{c\sigma}}, 
{\partial \over \partial A_{d\lambda}}, {\partial \over \partial A_{f\nu}}
\right\}, \quad
V = \left\{{\partial \over \partial u_{a}}, 
{\partial \over \partial A_{b\sigma}}, {\partial \over \partial A_{f\lambda}}, 
{\partial \over \partial \Phi_{c}}, {\partial \over \partial \Phi_{d}},
\right\},
\nonumber \\
V = \left\{{\partial \over \partial u_{a}}, 
{\partial \over \partial \Phi_{b}}, {\partial \over \partial \Phi_{c}}, 
{\partial \over \partial \Phi_{d}}, {\partial \over \partial \Phi_{e}},
\right\}
\eea

The first choice gives the Ward identity:
\bea
{\partial \over \partial x^{\mu}_{1}}
<\Omega, T\left({\partial^{2} \over \partial u_{a} \partial A_{f\nu}} 
T^{\mu}(x_{1}), 
{\partial \over \partial A_{b\rho}}T(x_{2}),
{\partial^{2} \over \partial A_{c\sigma} \partial A_{d\lambda}} T(x_{3})
\right)\Omega>
\nonumber \\
+ {\rm perm} (b\nu, c\sigma, d\lambda, f\nu)
+ (x_{1}, \leftrightarrow x_{2}) + (x_{1}, \leftrightarrow x_{3}) 
+ \cdots = 0
\eea
The chronological product involves the causal splitting of the following
commutator
\bea
\left[ {\partial^{2} \over \partial u_{a} \partial A_{f\nu}} T^{\mu}(x_{1}), 
T\left( {\partial \over \partial A_{b\rho}}T(x_{2}),
{\partial^{2} \over \partial A_{c\sigma} \partial A_{d\lambda}} T(x_{3})
\right) \right] 
\nonumber \\
= f_{afg} f_{gbd} f_{cde} 
(g^{\rho\lambda} g^{\nu\sigma} - g^{\rho\sigma} g^{\nu\lambda})
\partial^{\mu} D_{m_{g}}(x_{1}-x_{2}) \delta(x_{2}-x_{3}) + \cdots
\eea
which produces the anomaly
\be
A^{\rho\sigma\lambda\nu}_{abcde} = 2 f_{afg} f_{gbd} f_{cde} 
(g^{\rho\lambda} g^{\nu\sigma} - g^{\rho\sigma} g^{\nu\lambda})
\partial^{\mu} D_{m_{g}}(x_{1}-x_{2})
+ {\rm perm} (b\nu, c\sigma, d\lambda, f\nu) = 0. 
\ee

The other two cases can be treated similarly and do not produce anomalies. Let
us note that no finite renormalizations of the third order chronological
products are necessary to implement gauge invariance.

(iii) $n = 4,5$

In these cases, one can argue like in \cite{fermi} that only when the
chronological products involve at least one Fermionic loop one can have
anomalies. The relevant choices for the derivative set $V$ are similar to the
case (ii1) studied above. One obtains that the corresponding ward identities
might be broken by the box and the pentagon anomalies \cite{Ni1}. If this
anomalies are also zero, then we have gauge invariance up to the fifth order of
the perturbation theory, so according to the general theorem from the preceding
Section, we have gauge anomaly in all orders.


\newpage
\begin{thebibliography}{99}

\bibitem{AS}
A. Aste, G. Scharf,
``{\it Non-Abelian Gauge Theories as a Consequence of Perturbative Quantum
Gauge Invariance}",
hep-th/9803011, ZU-TH-6/98, Int. Journ. Mod. Phys. {\bf A 14} (1999) 3421-3432

\bibitem{ASD1}
A. Aste, G. Scharf, M. D\"utsch,
``{\it On Gauge Invariance and Spontaneous Symmetry Breaking},
J. Phys. {\bf A 30} (1997) 5785-5792

\bibitem{ASD3}
A. Aste, G. Scharf, M. D\"utsch,
``{\it Perturbative Gauge Invariance: the Electroweak Theory II},
hep-th/9702053, Ann Phys. (Leipzig) {\bf 8} (1999) 389-404

\bibitem{Bo}
F. M. Boas,
``{\it Gauge Theories in Local Causal Perturbation Theory}",
DESY-THESIS-1999-032, November 1999

\bibitem{BDF}
F. M. Boas, M. D\"utsch, K. Fredenhagen,
``{\it A Local (Perturbative) Construction of Observables in Gauge Theories:
Nonabelian Gauge Theories}", unpublished

\bibitem{BS2}
N. N. Bogoliubov, D. Shirkov,
``{\it Introduction to the Theory of Quantized Fields}",
John Wiley and Sons, 1976 (3rd edition) 

\bibitem{Du1}
M. D\"utsch,
``{\it On Gauge Invariance of Yang-Mills Theories with Matter Fields}",
Z\"urich-University-Preprint ZU-TH-10/95,
Il Nuovo Cimento {\bf A 109} (1996) 1145-1186

\bibitem{Du2}
M. D\"utsch,
``{\it Slavnov-Taylor Identities from the Causal Point of View}",
Z\"urich-University-Preprint ZU-TH-30/95, hep-th/9606105

\bibitem{Du3}
M. D\"utsch,
``{\it Non-Uniqueness of Quantized Yang-Mills Theories}",
Z\"urich-University-Preprint ZU-TH-9/96, hep-th/9606100,
J. Phys. {\bf A 29} (1996) 7597-7617

\bibitem{Du4}
M. D\"utsch,
``{\it Finite QED and Quantum Gauge Field Theory}",
Acta Phys. Polonica {\bf B 27} (1996) 2421-2440

\bibitem{DF1}
M. D\"utsch, K. Fredenhagen,
``{\it A Local (Perturbative) Construction of Observables in Gauge Theories:
the Example of QED}",
hep-th/9807078,
Commun. Math. Phys. {\bf 203} (1999) 71-105

\bibitem{DF3}
M. D\"utsch, K. Fredenhagen,
``{\it Algebraic Quantum Field Theory, Perturbation Theory, and the Loop
Expansion}",
hep-th/0007129, DESY-00-013

\bibitem{DG1}
J. Dimock, J. Glimm,
``{\it Measures on Schwartz Distribution Space and Applications to
$P(\phi)_{2}$ Field Theories}
Adv. Math. {\bf 12} (1974) 58

\bibitem{DHKS1}
M. D\"utsch, T. Hurth, F. Krahe, G. Scharf,
``{\it Causal Construction of Yang-Mills Theories. I}",
Il Nuovo Cimento {\bf A 106} (1993) 1029-1041

\bibitem{DHKS2}
M. D\"utsch, T. Hurth, F. Krahe, G. Scharf,
``{\it Causal Construction of Yang-Mills Theories. II}",
Il Nuovo Cimento {\bf A 107} (1994) 375-406

\bibitem{DHS1}
M. D\"utsch, T. Hurth, G. Scharf,
``{\it Gauge Invariance of Massless QED}",
Phys. Lett {\bf B 327} (1994) 166-170

\bibitem{DHS2}
M. D\"utsch, T. Hurth, G. Scharf,
``{\it Causal Construction of Yang-Mills Theories. III}",
Il Nuovo Cimento {\bf A 108} (1995) 679-707

\bibitem{DHS3}
M. D\"utsch, T. Hurth, G. Scharf,
``{\it Causal Construction of Yang-Mills Theories. IV. Unitarity}",
Il Nuovo Cimento {\bf A 108} (1995) 737-773

\bibitem{DS}
 M. D\"utsch, G. Scharf
``{\it Perturbative Gauge Invariance: the Electroweak Theory},
hep-th/9612091, Ann. Phys. (Leipzig) {\bf 8} (1999) 359-387

\bibitem{DS1}
 M. D\"utsch, B. Schroer,
``{\it Massive Vectormesons and Gauge Theory},
hep-th/9906089, preprint DESY 99-144, Journ. Phys. {\bf A 33} (2000) 4317-4356

\bibitem{EG1}
H. Epstein, V. Glaser,
``{\it The R\^ole of Locality in Perturbation Theory}",
Ann. Inst. H. Poincar\'e {\bf 19 A} (1973) 211-295

\bibitem{Gl}
V. Glaser,
``{\it Electrodynamique Quantique}",
L'enseignement du 3e cycle de la physique en Suisse Romande (CICP), Semestre
d'hiver 1972/73

\bibitem{YM} D. R. Grigore
``{\it On the Uniqueness of the Non-Abelian Gauge Theories in Epstein-Glaser 
Approach to Renormalisation Theory}", 
hep-th/9806244, Romanian J. Phys. {\bf 44} (1999) 853-913

\bibitem{standard} D. R. Grigore
``{\it The Standard Model and its Generalisations in Epstein-Glaser 
Approach to Renormalisation Theory}", 
hep-th/9810078, Journ. Phys. {\bf A 33} (2000) 8443-8476

\bibitem{fermi} D. R. Grigore
``{\it The Standard Model and its Generalisations in Epstein-Glaser 
Approach to Renormalisation Theory II: the Fermion Sector and the Axial Anomaly
}", 
hep-th/9903206, Journ. Phys. {\bf 34 A} (2001) 5429-5462

\bibitem{qed} D. R. Grigore,
``{\it Gauge Invariance of the Quantum Electrodynamics in the Causal Approach 
to Renormalization Theory}", 
hep-th/9911214, Ann. Phys. (Leipzig) {\bf 10} (2001) 439-471

\bibitem{scale} D. R. Grigore,
``{\it Scale Invariance in the Causal Approach to Renormalization Theory}", 
hep-th/0004163, Ann. Phys. (Leipzig) {\bf 10} (2001) 473-496

\bibitem{ren-ym} D. R. Grigore,
``{\it The Structure of the Anomalies of the Non-Abelian Gauge Theories 
in the Causal Approach }",
hep-th/0010226

\bibitem{wz} D. R. Grigore,
``{\it Wess-Zumino Model in the Causal Approach}",
hep-th/0011174, to appear in European Phys. Journ. {\bf C}

\bibitem{Gri1}
N. Grillo,
``{\it Some Aspects of Quantum Gravity in the Causal Approach}",
hep-th/9903011, the 4th workshop on ``Quantum Field Theory under the Influence
of External Conditions", Leipzig, Germany, Sept. 1998

\bibitem{Gri2}
N. Grillo,
``{\it Causal Quantum Gravity}",
hep-th/9910060

\bibitem{He1}
K. Hepp,
``{\it Renormalization Theory}",
in ``Statistical mechanics and Quantum Field Theory", Les Houches 1970,
Gordon \& Breach, N. Y.

\bibitem{Hu1}
T. Hurth,
``{\it Nonabelian Gauge Theories. The Causal Approach}",
Z\"urich-University-Preprint ZU-TH-36/94, hep-th/9511080,
Ann. of Phys. {\bf 244} (1995) 340-425

\bibitem{HS}
T. Hurth, K. Skenderis,
``{\it Quantum Noether Method}",
hep-th/9803030, Nucl. Phys. {\bf 541} (1999) 566-614

\bibitem{Kr1}
F. Krahe,
``{\it A Causal Approach to Massive Yang-Mills Theories}",
DIAS-STP-95-01

\bibitem{Ni1}
P. van Nieuwenhuizen,
``{\it Anomalies in Quantum Filed Theory: Cancellation of Anomalies in
$d = 10$ Supergravity}",
Lect. Notes in Math. and Theor. Phys., vol. 3, series B: Theor. Particle Phys.,
Leuven Univ. Press, 1988

\bibitem{Pr3}
D. Prange,
``{\it Energy Momentum Tensor and Operator Product Expansion in Local Causal
Perturbation Theory}", Ph. D. Thesis, 
hep-th/0009124

\bibitem{PS}
G. Popineau, R. Stora, ``{\it A Pedagogical Remark on the Main Theorem of
Perturbative Renormalization Theory}", unpublished preprint

\bibitem{Sc1}
G. Scharf,
``{\it Finite Quantum Electrodynamics: The Causal Approach}",
(second edition) Springer, 1995

\bibitem{Sc4}
G. Scharf,
``{\it General Massive Gauge Theory}",
Z\"urich Univ. preprint ZU-TH 1199, hep-th/9901140, 
Il Nuovo Cimento {\bf A 112} (1999) 619-638

\bibitem{Sc5}
G. Scharf,
``{\it Quantum Gauge Theories. A True Ghost Story}",
John Wiley, 2001

\bibitem{Sto1}
R. Stora,
``{\it Lagrangian Field Theory}",
Les Houches, 1971, C. De Witt, C. Itzykson eds.

\bibitem{St1}
O. Steinmann,
``{\it Perturbation Expansions in Axiomatic Field Theory}",
Lect. Notes in Phys. {\bf 11}, Springer, 1971

\bibitem{SW}
G. Scharf, M. Wellmann,
``{\it Quantum Gravity from Perturbative Gauge Invariance}",
hep-th/9903055

\bibitem{Va} 
V. S.Varadarajan, 
``{\it The Geometry of Quantum Theory}" (second edition),
Springer, 1985

\bibitem{WG}
A. S. Wightman, L. G\aa rding,
``{\it Fields as Operator-Valued Distributions in Relativistic Quantum Field
Theory}", Arkiv Fysik {\bf 28} (1965) 129-184
\end{thebibliography}
\end{document}